\documentclass[iop]{emulateapj}

\shortauthors{A. Cavaliere et al.}
\shorttitle{Cluster Grand Design}
\slugcomment{Accepted by ApJ}

\begin{document}

\title{A Grand Design for Galaxy Clusters: Connections and Predictions}
\author{A. Cavaliere\altaffilmark{1,2}, A. Lapi\altaffilmark{1,3}, R. Fusco-Femiano\altaffilmark{4}}
\altaffiltext{1}{Dip. Fisica, Univ. `Tor Vergata', Via Ricerca Scientifica 1,
00133 Roma, Italy.}\altaffiltext{2}{INAF, Osservatorio Astronomico di Roma,
via Frascati 33, 00040 Monteporzio, Italy} \altaffiltext{3}{SISSA, Via
Bonomea 265, 34136 Trieste, Italy.}\altaffiltext{4}{INAF-IASF, Via Fosso del
Cavaliere, 00133 Roma, Italy.}

\begin{abstract}
We take up from a library of $12$ galaxy clusters featuring extended X-ray
observations of their Intra Cluster Plasma (ICP), analyzed with our
entropy-based Supermodel. Its few intrinsic parameters $-$ basically, the
central level and the outer slope of the entropy profile $-$ enable us to
uniformly derive not only robust snapshots of the ICP thermal state, but also
the `concentration' parameter marking the age of the host dark matter halo.
We test these profiles for consistency with numerical simulations and
observations. We find the central and the outer entropy to correlate, so that
these clusters split into two main classes defined on the basis of low (LE)
or high (HE) entropy conditions prevailing \emph{throughout} the ICP. We also
find inverse \emph{correlations} between the central/outer entropy and the
halo concentration. We interpret these in terms of mapping the ICP progress
on timescales around $5$ Gyr toward higher concentrations, under the drive of
the dark matter halo development. The progress proceeds from HEs to LEs,
toward states of deeper entropy erosion by radiative cooling in the inner
regions, and of decreasing outer entropy production as the accretion peters
out. We propose these radial and time features to constitute a cluster Grand
Design, that we use here to derive a number of predictions. For HE clusters
we predict sustained outer temperature profiles. For LEs we expect the outer
entropy ramp to bend over, hence the temperature decline to steepen at low
$z$; this feature goes together with an increasing turbulent support, a
condition that can be directly probed with the SZ effect. We finally discuss
the looming out of two intermediate subsets: wiggled
$\mathrm{\widetilde{HE}}$ at low $z$ that feature central temperature
profiles retaining imprints of entropy discharged by AGNs or deep mergers;
high-$z$ LEs, where the cosmogony/cosmology had little time to
enforce a sharp outer entropy bending.
\end{abstract}


\section{Introduction}

\setcounter{footnote}{0}

Rich galaxy clusters, with their overall masses $M\sim 10^{15}\, M_{\odot}$
and large virial sizes $R\sim$ Mpcs\footnote{We adopt the standard
`concordance' cosmology (Komatsu et al. 2011), for which the virial radius
reads $R\approx R_{100}\approx 4\,R_{200}/3\approx 2\, R_{500}$ in terms of
the radii encircling an average overdensity that amounts to $100$, $200$ and
$500$, respectively, over the background density in the critical universe.},
constitute the most recent cosmic structures with high contrast, but still
developing at low redshifts. Their gravitationally dominant dark matter (DM)
halos contain an appreciable amount $m\approx 0.16\, M$ of hot, diffuse
baryons in the form of an intracluster plasma (ICP) at virial temperatures
$k_BT\sim G\, M\, m_p/10\, R\approx$ several keVs and with average densities
$n\sim 10^{-3}$ cm$^{-3}$. The ICP conditions can be probed in X-rays
through its strong bremsstrahlung emissions of powers $L_X\propto n^2\,
T^{1/2}\, R^3\approx 10^{44-45}$ erg s$^{-1}$.

Our main focus here will be on the physics of the ICP, and specifically on
its `entropy'
\begin{equation}
k\equiv k_B T/n^{2/3}~,
\end{equation}
or better adiabat (see Bower 1997), which is simply related to the true
specific entropy $s$ by $\Delta s\equiv 3/2\, \ln k$. The quantity $k$ will
conveniently constitute our leading state variable, due to its basic
properties: it is \emph{eroded} and eventually erased at the cluster center
by radiative cooling; it is \emph{produced} at shock fronts driven both by
supersonic inflows across the cluster boundary and by central outflows; it is
\emph{conserved} and stratified upon adiabatic compression of the outer
intergalactic medium (IGM) into the ICP contained by the DM potential well.

Our scope will be to relate the entropy levels in the ICP to the evolution of
the containing DM halos. As to the latter, we will refer to the standard
scenario including hierarchical formation and secondary infall, updated by
state-of-the-art $N-$body simulations and analytical works (e.g., Zhao et al.
2003, Fakhouri et al. 2010, Wang et al. 2011). This scenario envisages a
first stage of fast collapse and major mergers forming the halo bulk from the
top of the initial density perturbation; this is followed by a slow
development of the outskirts by accretion of diffuse matter and minor clumps
from the perturbation wings (details and further references are given in
Appendix A1). The two stages are separated by the redshift $z_t \approx
0.5-1$ when the circular velocity $v_R^2$ at the virial boundary attains its
maximal value; this epoch leaves a clear imprint in the halo `concentration'
parameter $c\equiv R/r_{-2}$ (the ratio of the virial radius to the reference
radius in the halo bulk where the DM density slope equals $-2$), that grows
after $z_t$ following $c(z)\approx 3.5\, (1+z_t)/(1+z)$.

After $z_t$ the halos attain a quasi-static equilibrium described by the
Jeans equation; the explicit solutions (`$\alpha$-profiles', with
$\alpha=1.27$ in rich clusters) for the density $\rho(r)$ and the
gravitational potential $\Phi(r)$ are given by Lapi \& Cavaliere (2009a,b)
and recalled in Eq.~(B3). Note that the physical scales including $r_{-2}$
are modulated by $c$.

We will fulfill our purpose with the use of two main tools: basic entropy
patterns, and the entropy-based equilibrium condition expressed by our
Supermodel.

\section{Entropy patterns}

The basic ICP entropy run we expect to apply throughout the ICP can
be rendered as a central level $k_c$ connecting to a rising ramp with slope
$a$ toward the outer value $k_2$ in the form (see Tozzi \& Norman 2001, Voit
2005)
\begin{equation}
k(r) = k_c + k_2\,(r/R)^a~,
\end{equation}
illustrated in Fig.~1 (red solid line). Next we discuss the physical origin
of such a minimally structured distribution.

\subsection{Central entropy}

In the \emph{central} range $r \la 2\times 10^2$ kpc the entropy is initially
set at levels $k_c\sim 10^2$ keV cm$^2$, not much exceeding the levels $k_1$
prevailing in the IGM (see Ryu et al. 2008, Nicastro et al. 2010). This is
because during the initial fast collapse the temperatures in the virialized
core are high, at $k_B\,T\approx G\,m_p\, M(<r)/10\,r \sim$ a few keVs, but
the ICP is dense at $n\sim 10^{-3}$ cm$^{-3}$, in step with the general
overdensities $\delta\,\rho/\rho \ga 2\times 10^2$ over the average
environment.

Such entropy levels are \emph{eroded} or even erased away following ${\rm
d}\,s/{\rm d}t = - s /t_c$, due to the radiative cooling by bremsstrahlung
(increasingly dominating over line emission for $k_B T\ga 2$ keV) that makes up
the observed X-ray emissions; the associated timescale for a single-phase ICP
(see Sarazin 1988) reads $t_c\approx 30\, (k_B T/{\rm keV})^{1/2}\,
(n/10^{-3}~{\rm cm}^{-3})^{-1}$ Gyr. Thus cooling may be slow and little
relevant in the low-density outskirts, but is speeded up in the dense central
ICP, so that within some $5$ Gyr the levels $k_c$ are depressed from $\sim
10^2$ keV cm$^2$ down to $\sim 10^1$ keV cm$^2$. Wherefrom cooling becomes so
fast as to match the dynamical times $\sim 10^{-1}$ Gyr, to the effect of
impairing the thermal pressure support; the process is even faster in
multi-phase ICP with a considerable cold component.

This leads to ICP condensation and to cooling faster yet, so starting up an
accelerated settling to the cluster center and onto the central galaxies (the
classic `cooling catastrophe'), were it not for renewed energy injections
(see Binney \& Tabor 1995, Cavaliere et al. 2002, Voit \& Donahue 2005,
Tucker et al. 2007, Hudson et al. 2010). These occur when the accretion
reaches down into the galactic nuclei and onto their central supermassive
black holes, to trigger or rekindle a loop of intermittent starbursts and AGN
activity; in the form of gentle bubbling or moderate outbursts over some
$10^{-1}$ Gyr, this can stabilize the time-integrated $k_c$ at levels of
$\sim 10^1$ keV cm$^2$. Such an enticing scenario is discussed, among others,
by Ciotti \& Ostriker (2007), McNamara \& Nulsen (2007), Conroy \& Ostriker
(2008), Churazov (2010). In sum, a cool core constitutes an attractor for the
thermal state of the central ICP.

On the other hand, $k_c$ may be \emph{raised} up to levels of several $10^2$
keV cm$^2$ when substantial energy injections $\Delta E$ occur into the ICP
from violent outbursts of AGNs in central galaxies or even more from deep
mergers. These injections drive through the central ICP a blastwave bounded
by a leading shock with Mach number gauged by the relation $\mathcal{M}^2
\approx 1 + \Delta E/E$ in terms of the ICP thermal energy $E\approx 2\times
10^{61}\, (k_BT/\mathrm{keV})^{5/2}$ erg (see Lapi et al. 2005, their
Fig.~7); a strong shock with $\mathcal{M}^2\ga 3$ would require injections
$\Delta E\ga 2\, E$, i.e., a few tens of keVs/particle. This may be the case
for deep major mergers, while it is hardly matched by an AGN powered by a
supermassive black hole up to $5\times 10^{9}\, M_{\odot}$ with only some
$5\%$ of the energy discharged effectively coupled to the ICP (see Lapi et
al. 2005).

Blasts that preserve the overall equilibrium may still leave a long-lasting
imprint onto the central ICP in the form of an entropy addition spread out to
a radius $r_f\approx 10^2$ kpc where the blast has expanded, stalled and
degraded into sound waves. A handy representation (see Fusco-Femiano et al.
2009; see also Appendix B) of the entropy distribution in such conditions is
still given by Eq.~(2) for $r\geq r_f$, while a roughly constant level
\begin{equation}
k(r)=k_c
\end{equation}
applies for $r\leq r_f$, as illustrated in Fig.~1 (orange dashed line). We
shall see that conspicuous central wiggles may then appear in the radial
temperature profiles if these imprints survive spherical averaging; such
features will persist over timescales longer than the blast transit time
$\sim 0.3$ Gyr if shorter than the cooling time $\sim 5$ Gyr.

Stronger if rarer energy injections with $\Delta E\gg E$ can be produced as
head-on major mergers following the halo bulk collapse (see McCarthy et al.
2007, Norman 2011) deposit at the center large energies of several tens keV
per particle, and entropy levels $\gg 10^2$ keV cm$^2$; these trigger
conditions of severe disequilibrium such as in A754 (see Macario et al. 2011)
and A2146 (see Russell et al. 2010), or outright disruption like in MACS
J0025.4-1222 (see Brada\v{c} et al. 2008), or 1E0657-56 (the `Bullet
Cluster', see Clowe et al. 2006).

\subsection{Outer entropy}

Supersonic inflows of external IGM drive at the cluster \emph{boundary}
$R\sim$ Mpc strong shocks intertwined into a web or layer located at
$R_s\approx R$ where accretion feeds on filaments (see Lapi et al. 2005, Voit
2005). These shocks are effective in thermalizing a considerable fraction of
the specific energy $v^2_1$ that the IGM gains as it infalls from the
`turnaround' radius $R_{\rm ta}$ to the virial $R\approx R_{\rm ta}/2$, to
the effect of providing substantial temperature jumps $T_2/T_1$ from the IGM
values. These jumps grow with the Mach number squared $\mathcal{M}^{2}$,
whilst the density jumps $n_2/n_1$ saturate to $4$, and the post-shock
kinetic energy $v^2_2/v^2_1$ decreases with $\mathcal{M}^{-2}$ in the shock
restframe, cf. Appendix A and Fig.~A2. As a result, soon after the cluster
formation large thermal energies are deposited in the thin ICP at the
boundary, with densities close still to the IGM's; there the entropy levels
reach up to $k_2\sim 5\times 10^3$ keV cm$^2$.

The bearing of these issues to the ICP physics is focused from the expression
derived by Cavaliere et al. (2009) for the value of the entropy slope $a_R$
at the boundary
\begin{equation}
a_R = 2.5-0.5\,b_R~.
\end{equation}
This value (clearly smaller than $2.5$) sensitively depends on the ratio
$b_R\equiv\mu m_p\, v_R^2/k_B T_2\ga 1$ of the potential to the ICP thermal
energy at $r=R$ (see Lapi et al. 2005). Values $a_R\approx 1.1$ are obtained soon
after the bulk collapse, when the inflow is still sustained and strong shocks
fully thermalize the infall energy $v_1^2=2\,\Delta \Phi$ into three degrees
of freedom, and produce postshock temperatures $k_B T_2\approx \mu m_p\,
v_1^2/3$ (for closer evaluations see Appendix A3-A4). On expressing the
potential drop from the turnaround to the shock in the form
$\Delta\Phi/v_R^2\approx 0.57$ (see Cavaliere et al. 2009), the standard
values $b_R\approx 3\, v_R^2/2\,\Delta\Phi\approx 2.7$ and $a\approx 1.1$
are obtained (Tozzi \& Norman 2001).

Eq.~(4) is derived as the current boundary value $a_R$ for $a$, but it
clearly yields also the running slope $a(r)$ in the middle range on
considering that $-$ in the absence there of energy sources $-$ the entropy
will be conserved and \emph{stratified} at the values previously produced
when the boundary was just at $r$. In other words, the radial entropy
distribution preserves the memory of the past-time development.

As the cluster outskirts grow farther out, the inflows slow down
considerably, and do so especially at low $z$; this straightforwardly is to
occur when the accretion is drawn from the tapering wings of a DM
perturbation over a background lowering under the accelerated cosmic
expansion. Thus the potential drop $\Delta\Phi$ becomes shallower (see
Appendix A1; also Lapi et al. 2010) while the shocks outgrow $R$, to the
effect of weakening the shock jumps and lowering $T_2$ toward the external
value $T_1$. As a result, $b_R$ grows and $a$ decreases toward zero.

A handy representation (see Lapi et al. 2010; see also Appendix B) of the
ensuing entropy distribution is still given by Eq.~(2) inside $r\leq r_b$
with $r_b\sim R/3$ (to be discussed in \S~4), while
\begin{equation}
k(r)=k_2\, (r/R)^{a+a'}\,e^{a'\,(R-r)/r_b}
\end{equation}
applies for $r>r_b$, as illustrated in Fig.~1 (blue dotted line). This
expression describes a simple linear decline of the slope $a(r)$ with a
gradient $a'\equiv (a-a_R)/(R/r_b-1)$ from the inner value $a\sim 1.1$ to the
outer value $a_R<a$.

Such an entropy \emph{bending} takes place on the timescale set by the
outskirts development, when the DM halo grows its concentration to values
$c\ga 6$ from the initial values $c\approx 3.5$ set soon after the bulk
collapse at $z_t$; e.g., for a cluster collapsed at $z_t\approx 1$ and
observed at $z\approx 0.15$ the time elapsed amounts to $6$ Gyrs.

In sum, the outer ramp flattens and bends over a timescale of several Gyrs,
while the central level $k_c$ is eroded away by radiative cooling. These two
changes are independently driven at far apart locations by quite different
processes; what they have in common, though, is their progress in time. So
one expectation from our picture is that they should take place
\emph{together} as the structures age, a main feature in our cluster
classification of \S~5.

\section{The entropy-based equilibrium}

The entropy-based equilibrium of the ICP within the DM gravitational wells is
constituted by our Supermodel (see Cavaliere et al. 2009 and Fusco-Femiano et
al. 2009), with the related straight algebra recapped in Appendix B.

\subsection{Thermal support}

There we recall that the linked radial profiles of temperature and density
read
\begin{equation}
k_B T(r)= n^{2/3}(r)\,k(r)\propto k^{3/5}(r)\,
\left[1+2/5\,b_R\,\mathcal{I}(r)\right]~,
\end{equation}
having used the shorthand $\mathcal{I}(r)\equiv \int_{r/R}^1{\rm d}x\,
[v_c^2(x)/v_R^2]\, [k(x)/k_2]^{-3/5}/x$ in terms of the circular velocity
$v_c^2\equiv G\, M/R$ (see also Appendix B, below Eq.~B3). As discussed by
Cavaliere et al. (2009), in the outskirts $\mathcal{I}$ is small compared to
$1$ and the whole factor in square parenthesis behaves like
$(r/R)^{-2\,b_R/5}$; on the other hand, at the center the integral
$\mathcal{I}\propto k_c^{-1/4}$ dominates over $1$ and scales inversely with
the central entropy level $k_c$.

These temperature and density profiles provide the volume emissivity for
bremsstrahlung, proportional to
\begin{equation}
S_X\propto n^2(r)\, T^{1/2}(r)\propto k^{-9/10}(r)\,
\left[1+2/5\,b_R\,\mathcal{I}(r)\right]^{7/2}~;
\end{equation}
this constitutes the basis for computing (after spectral-bandpass windowing
and projection) the X-ray observables, namely, the surface brightness
and the emission-weighted temperature; full expressions are given in Appendix
B.

We stress that all these profiles for $n(r)$, $T(r)$, and $S_X(r)$ are
\emph{linked} together by the underlying entropy distribution. For a relevant
example, Eq.~(6) yields the central scaling laws $T_c\propto k_c^{0.35}$ and
$n_c^2\, T_c^{1/2}\propto k_c^{-1.8}$, see Cavaliere et al. (2009); thus when
$k_c$ is \emph{low} the temperature will \emph{dip} and the associated
emissivity will \emph{rise} toward the center, features that constitute the
marks of the conventional cool-core designation. On the other hand,
\emph{high} $k_c$ produce \emph{flat} emissivity profiles together with a
wide temperature \emph{plateau}, typical of the conventional non-cool-core
designation. Moreover, the central cooling time in single-phase equilibrium
may be expressed in terms of the entropy level $k_c$ only, to read simply
$t_c\approx 0.5\,(k_c/15\,{\rm keV~cm}^2)^{1.2}$ Gyr; this implies that high
levels of $k_c\approx 150$ keV cm$^2$ require timescale of order $8$ Gyr to
be eroded. In the outskirts, instead, the scaling $T(r)\propto r^{7/5\,
a_R-2}$ holds, showing that when $k(r)$ is \emph{bent} down with $a_R\ll 1$ the
temperature will fall steeply outwards; in simple terms, the profile
$T(r)\propto k(r)\, n^{2/3}(r)$ will follow $n^{2/3}(r)$ or steeper when
$k(r)$ is nearly constant or even bent down. Meanwhile, the brightness will be
flatter at intermediate radii (see Fig.~2), and constitutes a simple pointer
toward interesting temperature and entropy distributions (see Cavaliere et
al. 2011).

An observable independent of X-ray data is provided by the SZ effect (Sunyaev
\& Zel'dovich 1972); the radial profile of its strength parameter is
proportional to the thermal ICP pressure and writes as
\begin{equation}
y(r)\propto n(r)\, T(r) \propto
\left[1+2/5\,b_R\,\mathcal{I}(r)\right]^{5/2}~.
\end{equation}
We stress that the Supermodel implies a nearly universal pressure profile
(and correspondingly for the SZ effect), since the entropy radial dependence
is encased into the slowly varying factor $\mathcal{I}(r)$; this is the
ultimate origin for the approximate invariance of the pressure profile
derived from the X-ray data by Arnaud et al. (2010). Using the inner scaling
of $n$ and $T$ with $k_c$ we find that the scaling $y\propto k_c^{-0.65}$
holds, implying that higher values of $y$ correspond to lower $k_c$. At the
other end, in the outskirts $y\propto r^{2\, a_R-5}$ applies, implying
sharper declines in clusters with shallower entropy ramps. Thus the
(projected) SZ effect provides a direct probe of the entropy levels
throughout a cluster, and so an independent way for classifying HE and LE
types from \textsl{Planck} (see Aghanim et al. 2011) and from ground-based
instruments.

\subsection{Turbulent support}

As argued above, the conditions of low entropy production are related to
mildly supersonic inflows and weak boundary shocks with decreasing Mach
number $\mathcal{M}^2<3$; we stress that in turn they are conducive to trigger
outer subsonic turbulence developing under the drive of relatively more
inflow energy $v_2^2/v_1^2\propto \mathcal{M}^{-2}$ seeping through the
weaker shocks (see Cavaliere et al. 2011; also Appendix A4 for details). The
turbulent contribution to equilibrium is conveniently described in terms of
the ratio $\delta \equiv p_{\rm nth}/p_{\rm th}$ of the turbulent to thermal
pressure. The boundary normalization is consistently set by $\delta_R\propto
v_2^2/v_1^2$, while the shape $\delta(r)/\delta_R$ of its inward decline on a
scale $\ell\sim 10^2$ kpc is provided by the classic cascade from large
`eddies' at the macroscopic coherence length, fragmenting to small eddies
where dissipation becomes effective (see Kolmogorov 1941, Monin \& Yaglom
1965; see \S~B3 in Appendix B for details).

In fact, it turns out that the total pressure $p_{\rm th}+p_{\rm nth}=p_{\rm
th}\, (1+\delta)$ can be straightforwardly included in the hydrostatic
equilibrium solved by the Supermodel; the result can be described simply in
terms of Eq.~(6), with $T$ and $k$ replaced everywhere (including
$\mathcal{I}$) by $\tilde T\equiv T\,(1+\delta)$ and by
\begin{equation}
\tilde k\equiv k\,(1+\delta)~.
\end{equation}
The underlying rationale is that turbulent eddies add to the microscopic
thermal degrees of freedom in dispersing and ultimately dissipating the
inflow kinetic energy $v_2^2$ seeped through the shock.

While turbulence is stirred, the thermal pressure required for overall
support in the given DM gravitational potential well is decreased. If
turbulence were not accounted for, the overall masses estimated from X rays
would tend to be negatively biased compared to the gravitational lensing
measurements (Nagai et al. 2007, Lau et al. 2009, Meneghetti et al. 2010,
Kawaharada et al. 2010). Meanwhile, the intensity parameter of the volume
thermal SZ effect $y(r)$ is lowered relative to the pure thermal equilibrium
expression Eq.~(8) by an explicit factor $1/(1+\delta)$, adding to small
corrections to the integrand inside $\mathcal{I}$. Note that such a
straightforward lowering is considerably stronger than may result from any
reasonable ion-electron disequilibrium at the shock (see the accurate
estimates by Wong \& Sarazin 2009). Thus SZ effect can also provide a
\emph{direct} probe of a low thermal pressure, which implies a considerable
turbulent component in the cluster outskirts for attaining equilibrium (see
Cavaliere et al. 2011). The dearth of outer thermal pressure is indicated by
stacked \textsl{WMAP} data (see Komatsu et al. 2011); the contribution to
such conditions from LEs and HEs will be discussed in a forthcoming paper.

\bigskip

\section{Toward a cluster library}

We aim at constructing first a library of clusters from extended
circularly-averaged data; on this basis we aim at introducing a physically
meaningful cluster classification scheme and at discussing the connection of
the ICP thermal state to the DM halo development. This requires robust fits
to the X-ray observables from linked, \emph{consistent} profiles of density
and temperature, as to pinpoint the few independent parameters governing the
ICP entropy distribution. Specifically, we adopt the following strategy.

The entropy-based picture of \S~2 (illustrated in Fig.~1) suggests the basic
entropy distribution of Eq.~(2), constituted by a central level $k_c$ going
into a ramp rising with slope $a\approx 1$ toward the outer value $k_2$. Two
relevant and alternative variants may apply: the central floor $k_c$ extends
out to a radius $r_f$ and is angled there to the outer ramp, see Eq.~(3);
beyond a radius $r_b$ the ramp bends over to a shallow slope $a_R\ll 1$ joining
a low boundary value $k_2$, see Eq.~(5). The basic distribution provides a
baseline with a minimal number of parameters; the first variant is convenient
when central temperatures are high, but wiggles stand out, while the second
applies to cases with low central temperature and steep temperature decline
into the outer region.

We first insert the basic entropy distribution Eq.~(2) with free shape
parameters $k_c$ and $a_R$ into the Supermodel Eq.~(6), and derive the radial
profiles of density and temperature; the Supermodel algorithm is made
available at the URL \texttt{http://people.sissa.it/ $\sim$lapi/Supermodel/}.
We then perform a fit to the projected, emission-weighted temperature data,
using a multiparametric $\chi^2$ minimization procedure (e.g., \textsl{MPFIT}
by Markwardt 2009), and derive the temperature scale $T_2$ at the boundary
(see Eq.~B7). Finally, we fit the surface brightness including the bandpass
correction (that requires $T_2$), and derive the scale $n_2$ at the boundary
(see Eq.~B8). Thus we obtain also the entropy normalization $k_2=k_B
T_2/n_2^{2/3}$ at the boundary, to complete the entropy distribution.

When the $\chi^2$ value of a fit turns out to be large, we proceed to insert
in the Supermodel the variant entropy distributions given by either Eq.~(3)
or Eq.~(5); this adds a further parameter, either $r_f$ (for A644 and A2256)
or $r_b$ (for A1795, PKS0745, A2204, A1413). The statistical significance of
the added parameter is corroborated by relevant improvements in the reduced
$\chi^2$ values (see Table~1). This is further checked with the $F-$test,
which yields a significance level larger than $98\%$ but for A1413 where it
is larger than $96\%$.

We stress that our Supermodel fits are performed over the whole radial range
covered by the current X-ray data. In a number of clusters observed by
\textsl{Suzaku} (e.g., A1795), the X-ray data extend out to approach the
virial radius $R$; in other instances observed by \textsl{XMM-Newton} (e.g.,
A1656) the data are more limited (around $R_{500}$), and the outer parameters
are provided by the Supermodel upon extrapolation, implying larger
uncertainties. In the case of PKS0745 the uncertainty is particularly large
due to discrepancies between the \textsl{XMM-Newton} and \textsl{Suzaku}
datasets, see the discussion by Eckert et al. (2011).

We note that the shape parameters $k_c$ and $a_R$ may be determined from
fitting either the temperature or the brightness profile; the results are
consistent within the respective uncertainties, but the value derived from
the former is to be preferred whenever extended, high-quality data are
available (with the due caveats discussed by Eckert et al. 2011 as to
anomalous background, and by Simionescu et al. 2011 as to effects of possible
clumpiness in one sector of the Perseus cluster), since the temperature
dominates the brightness in the entropy expression $k\propto T^{7/6}\,
S_X^{-1/3}$ (see Cavaliere et al. 2005).

On the other hand, fits to the X-ray brightness can also provide the DM
concentration $c=R/r_{-2}$ that enters the Supermodel formalism through
$v_c^2(r)$, while the outer scale $R$ is provided by independent observations
such as the red-sequence termination or gravitational lensing. We note that
the determination of $c$ is mainly based on the outer brightness data, so is
closely independent of the inner entropy distribution. We stress that the
Supermodel leads to a fast yet robust evaluation of $c$ from X rays only,
with results less biased than gravitational lensing by prolateness effects
(discussed, e.g., by Corless et al. 2009).

We also note that the few parameters entering the entropy distribution are
calibrated from fitting with the Supermodel the observables directly
expressed in terms of the radial profiles of $n(r)$ and $T(r)$, with no need
for delicate data deprojections (discussed by Yoshikawa \& Suto 1999,
Cavaliere et al. 2005, Croston et al. 2006, Urban et al. 2011).

The ICP parameters so derived are used here to build up the library of $12$
clusters presented in Table~1. Ten of these have been analyzed by us in
previous works (Cavaliere et al. 2009, Fusco-Femiano et al. 2009, Lapi et al.
2010), while here we add A399 and refine the analysis of A2218. Examples of
Supermodel fits are illustrated in Fig.~2.

One may ask to what extent the entropy distributions derived from the
Supermodel might depend on the underlying assumptions. These concerns are
swept away by Fig.~3 where we compare, in the radial range $r\ga 0.2\, R$
where cooling is negligible, the entropy distributions derived from our
Supermodel analysis of the $12$ clusters listed in Table~1, with the outcomes
of the nonradiative hydro-simulations by Burns et al. (2010) for a sample of
$24$ relaxed massive clusters. Our results are seen to be consistent with the
simulation outcomes and their variance, that grows wider into the outskirts.

Moreover, such comparison implies that throughout most of the cluster volume
the Supermodel results are robust against the assumptions of spherical
symmetry, hydrostatic equilibrium, and purely smooth accretion. In fact, in
the inner regions merger-related geometrical asymmetries are smoothed out on
a crossing timescale, shorter than the time required by cooling to erase
entropy excesses of $\sim 10^2$ keV cm$^2$. In the middle regions,
approximately spherical symmetry of the ICP is indicated by various
simulations (e.g., Lau et al. 2011). In the outer regions, the accretion is
dominated by minor mergers or truly diffuse matter, as shown in detail by the
simulations of Wang et al. (2011, see their Fig.~7). All that explains why
the snapshots provided by our Supermodel fits to the X-ray data are so
robust.

\bigskip
\bigskip

\section{A cluster Grand Design}

We have parted the $12$ clusters listed in Table 1 into two main blocks on
the basis of their $k_c$ values being of the order of a few $10^1$ or a few
$10^2$ keV cm$^2$; within each block, we have ordered the clusters on the
basis of their $a_R$ values. It is easily perceived that the two main blocks
are also parted as to the values of their DM concentration $c$. The ordering
indicates \emph{correlations} between these basic physical parameters
quantified in \S~5.1, and suggests an evolutionary \emph{trend} linking the
ICP thermal state with the DM development to be discussed in \S~5.2.

\subsection{Correlations}

In the top panel of Fig.~4 we illustrate the central entropy level $k_c$ vs.
the entropy slope $a$ in the cluster bulk at $R_{500}\approx 0.5\, R$. For
cool-core clusters our results (blue dots) from the Supermodel analysis
compare well as to central values and their uncertainties with the sample of
relaxed, mostly cool-core clusters by Pratt et al. (2010; green squares). It
is seen that as to the average data values, $k_c$ correlates poorly with $a$
at $R_{500}$, as quantified by the low value of the Spearman's rank
correlation coefficient $\rho\approx 0.27$ (cf. Lupton 1993) both for our and
the above authors' samples.

In the middle panel of Fig.~4 we illustrate the central levels $k_c$ vs. the
outer slopes $a_R$. As to the latter we find values close to $a$ for
non-cool-core clusters (red dots), while appreciably lower for cool-core
clusters (blue dots). It is seen that $a_R$ correlates well with $k_c$ as to
the average values; this is quantified by the value of the Spearman's rank
correlation coefficient $\rho\approx 0.64$. On the other hand, the often
large uncertainties in $a_R$ (related to true uncertainties in the outer
X-ray data) and especially in $k_c$ (related also to inner physical
complexities) will blur the correlation.

We test to what degree this occurs by running $10^5$ Monte Carlo simulations,
randomly sampling values of $k_c$ and $a_R$ from Gaussian distributions
around their averages, with widths given by their formal $1-\sigma$
uncertainties in both variables; with this conservative treatment the average
Spearman's coefficient is lowered to $\rho\approx 0.44$, corresponding to a
$9\%$ probability for chance occurrence of the correlation. In addition, we
compute that on statistical grounds the probability of `outliers' (objects
with $k_c\geq 30$ keV cm$^2$ and $a_R\leq 0.6$) is $5\%$ on average, with
a formal standard deviation of $22\%$; this implies that on doubling the size
of present sample to $24$ objects, one should expect from $1$ to $7$
outliers. The above outcomes motivate us to investigate in \S~5.2 whether a
physical basis underlies the apparent dearth of clusters with high central
entropy levels $k_c>10^2$ keV cm$^2$ and low outer entropy production
$a_R<1$.

In the bottom panel of Fig.~4 we illustrate the outer slopes $a_R$ vs. the
concentration parameter $c$, derived with the Supermodel. We find that low
values of $a_R$ correspond to high values of $c$, which mark a long lifetime
from the formation $z_t$ to the observation redshift $z\approx 0$ following
$c\approx 3.5\,(1+z_t)/(1+z)$. Such an anti-correlation between $a_R$ and $c$
is highly significant as for the average data values, with a Spearman's
coefficient $\rho\approx -0.79$; on the other hand, the additional
uncertainties that also affect $c$ lower it to the conservative value
$\rho\approx -0.46$. Again, this outcome stimulates us to investigate any
physical dearth of clusters with high concentration $c>6$ and steep outer
slopes $a_R>1$, and offers a pattern to be confronted with future real and/or
virtual datasets.

\subsection{Classes}

Guided by the above discussion, we submit that all our clusters may be parted
into two main classes, defined on the basis of low entropy (LE) or high
entropy (HE) prevailing not only in the inner region but also
\emph{throughout} the ICP.

$\bullet$ HE clusters, featuring \emph{high} entropy \emph{throughout} the
ICP; that is, featuring not only a central level $k_c\approx 3\times 10^2$
keV cm$^2$, but also a very high boundary level $k_2\approx 3-5\times 10^3$
keV cm$^2$ corresponding to a steep entropy ramp with $a\ga 1$ throughout the
outskirts. The high values of $k_c$ yield a monotonic temperature profile
$T(r)$ throughout, slowly declining from the central plateau into the
outskirts. We stress that our class definition includes not only a central
non-cool-core state as in the designation introduced by Molendi \& Pizzolato
(2001) and pursued by Leccardi et al. (2010), but also an associated high
level of outer entropy production. We propose that the association arises due
to the \emph{young} age of the containing DM halos, marked by low values of
the concentrations $c\approx 4$, with a lifetime too short for central
entropy to be erased away and any entropy bending to be effective in the
outskirts.

$\bullet$ LE clusters, featuring low entropy \emph{throughout} the ICP; this
includes both a \emph{low} central baseline $k_c<30$ keV cm$^2$ and a
moderate outer level $k_2\la 10^3$ keV cm$^2$, so as to imply a ramp bending
outwards of $r_b/R\ga 0.3$ toward $a(r)<1$ (see also Hoshino et al. 2010);
the outcome is a low central value of $T$ and a \emph{peak} of $T(r)$ at
$r/R\la 0.2$ followed by a steep decline outwards, particularly effective at
low $z$ (e.g., A1795). Our class definition includes not only a central
cool-core state as in the standard designation, but also an associated low
level of outer entropy production. We propose that the association low $k_c$
$-$ shallow $a_R$ is to be traced back to the long lifetime of the containing
DM halos, marked by \emph{high} values of the concentrations $c\ga 6$. We
relate such a late stage in the outskirts development to dwindling inflows
that cause weaker boundary shocks with $\mathcal{M}^2\la 3$, low entropy
production and a substantial fraction of kinetic energy left over to drive
outer turbulent eddies.

The low $k_c$ levels proper to LEs are driven by cooling timescales $t_c$
shorter than the halo dynamical age marked by $c$. In fact, the divide
between LEs and HEs is around $k_c\approx 150$ keV cm$^2$ corresponding to a
cooling time $t_c\approx 8$ Gyr (e.g., the lapse between $z\approx 1$ and
$z\approx 0.1$); after this, fast cooling leads to an accelerated progress
toward $k_c$ levels lower yet. Eventually, however, the levels of $k_c$ are
likely to be stabilized by two additional physical processes, i.e.,
intermittent AGN activity and impacts of deep major mergers; two modes are
suggested by the broad, possibly double-peaked distribution for the number of
clusters with given $k_c$, as observed by Cavagnolo et al. (2009) and Pratt
et al. (2010), and discussed by Cavaliere et al. (2009).

The relationship between the classes is illustrated in the
\emph{evolutionary} chart of Fig.~6, that represents our cluster Grand
Design. This envisages clusters mainly born in a HE state of high entropy,
dominated by the fast violent collapse of the halo bulk and related strong
shocks in the infalling gas. Subsequently, on a timescale of several Gyrs
they progress toward a LE state since both the central entropy is lowered by
radiative cooling, and the outer entropy bends over because of the weakened
shocks and tapering entropy production. The Grand Design envisages that in a
number of cases such a sequence may be halted within a few Gyrs and
\emph{reversed} by late, trailing deep mergers which remold any nascent
cool-core and \emph{rejuvenate} the central ICP into a higher entropy state.

In fact, these clusters with ICP lingering in such an intermediate state may
be conveniently ranked in a subclass labeled $\mathrm{\widetilde{HE}}$,
marked out from basic HE by a wiggled central temperature profile. We have
shown (see discussion by Fusco-Femiano et al. 2009) that such profiles obtain
whenever the central entropy features a floor $k_c$ extended out to a radius
$r_f\sim 10^2$ kpc; correspondingly, the central brightness features a
particularly flat profile. We recall from \S~2 that such entropy additions
are likely imprinted by a blastwave with Mach numbers $\mathcal{M}^2\ga 3$
launched outwards by a head-on impact of a deep merger. When the blast has
stalled around $r_f$ and the overall equilibrium in the ICP is recovered, the
central entropy is still enhanced to levels up to $k_c\sim 10^2$ keV cm$^2$,
and so is immune to subsequent, weaker AGN-driven blasts. Such a
$\mathrm{\widetilde{HE}}$ morphology turns out to occur not only in the two
cases listed in our Table~1, but also in several more instances of the kind
illustrated by Rossetti \& Molendi (2010), close to $50\%$ of their
non-cool-core clusters; thus we propose the $\mathrm{\widetilde{HE}}$s to
deserve a subclass status.

Our interpretation of the $\mathrm{\widetilde{HE}}$ morphology relates the
size $r_f$ to the epoch of the merger responsible for the entropy input; such
an epoch is expected to be in between the blast transit time
$r_f/\mathcal{M}\, c_s\sim$ some $10^{-1}$ Gyr and the several Gyrs taken by
radiation to erode the floor, or by central turbulence to blur it. Such a
timing ensures an accurate description of the ICP thermodynamics by the
Supermodel based on hydrostatic equilibrium. Note also that the ICP attains
its equilibrium somewhat faster than the DM does (see Ricker \& Sarazin 2001,
Lapi et al. 2005), while the circularized data (see Snowden et al. 2008) tend
to smooth out limited deviations from spherical hydrostatics (for a detailed
discussion see Fusco-Femiano et al. 2009).

\section{Predictions from the Grand Design}

Here we present other, specific predictions derived from our cluster Grand
Design.

$\bullet$ We expect the HE clusters to feature outer profiles, $T(r)$
declining mildly to a boundary value $T_2$ still \emph{sustained}. We
illustrate in Fig.~5 our prediction for such a mild decline in the two HE
clusters A1656 and A2256, compared with the currently limited data. Note that
for A1656 our Supermodel fit has only used the data by Snowden et al. (2008)
out to $r\approx R/3$; our outer prediction agrees with the recent data by
Wik et al. (2009) extending out to $R/2$.

$\bullet$ We expect LE clusters to feature at \emph{low} $z$ particularly
\emph{small} values of $k_c$ and \emph{sharply} bent outer entropy profiles.
The latter yield \emph{steeply} declining $T(r)$ profiles, as supported by
the \textsl{Suzaku} observations of a few clusters like A1795; a similar
recent case may be constituted by A2142, see Akamatsu et al. (2011). Low SZ
signals and increasing contribution of outer turbulent support are also
expected; relatedly, in these clusters the mass reconstructed from X-ray
observations will show systematic \emph{deficits} relative to the
gravitational lensing result (see Cavaliere et al. 2011).

$\bullet$ We expect at \emph{higher} $z$ a \emph{lower} fraction of LEs,
reflecting the main evolutionary trend from HEs to LEs envisaged by our Grand
Design; this is consistent with the evidence by Santos et al. (2010) based on
observing the average surface brightness up to redshift $z\approx 1.3$. When
observations of very low surface brightness will become feasible, we expect
steeper brightness profiles and a milder temperature decline in the outskirts
to loom out (see \S~3.1), as for such high-$z$ LEs the cosmology/cosmogony
had not time enough to sharpen the outer entropy bending.

Our picture envisaging low or high entropy levels to hold throughout the ICP
is consistent with the present dearth of the following pairings: nearby
clusters with low $k_c$ levels and high $a_R$ values (that would be located
in the upper left strip of the $k_c-a_R$ plane in Fig.~4); clusters with high
$k_c>10^2$ keV cm$^2$ and low $a_R<1$ (that would be located in the lower
right corner of the $k_c-a_R$ plane in Fig.~4); inner temperature wiggles in
highly concentrated clusters with $c\ga 6$. Wider libraries based on
extended, high-quality temperature data will allow testing the above
predictions. We add that the Supermodel predicts the projected SZ effects
(otherwise closely universal, \S~3.1) to differ from the HE to the LE cluster
population, with the latter featuring steeper profiles in the outskirts; in
fact, as stated in \S~3.2 in LEs outer turbulence is expected to contribute
substantially to the equilibrium, so as to lower by $1+\delta$ the thermal SZ
effect.

\section{Discussion and Conclusions}

We have seen how the cluster thermodynamical state of nearly relaxed clusters
can be probed by means of \emph{linked}, robust profiles of density and
temperature in the ICP derived from extended X-ray data. We have carried out
such a task on using the Supermodel formalism with its few, \emph{intrinsic}
parameters that modulate the underlying distribution of the specific entropy.
On this basis, we have grouped the rich clusters analyzed here into two main
classes, on account of low (LE) or high entropy (HE) prevailing
\emph{throughout} the ICP.

Such classes constitute thermal conditions with long persistence. In fact,
HEs with their hot atmospheres are stabilized by long cooling times and
stubborn resistance to supersonic flows; LEs are likely stabilized by inner
AGN energy injections, while their outskirts evolve slowly as the inflows
across the boundary decrease toward low $z$. The main overall evolutionary
course proceeds from HEs to LEs due to erosion of central entropy by cooling,
and to reduced production of outer entropy by weakened accretion shocks.

However, such a course may be interrupted or even reversed by large entropy
injections from major mergers, particularly frequent at high $z$; thus an
$\mathrm{\widetilde{HE}}$ thermal state sets in, marked out by wiggles in the
central radial temperature distribution. These are interpreted in the
Supermodel framework in terms of a sharp entropy floor extending out to
$r_f\ga 50$ kpc and with levels $k_c$ around $200$ keV cm$^2$. We consider
these as intermediate objects, constituting a subclass contiguous to, and
blending into the HE main class.

Our overall picture derived from the snapshots condensed in Table~1
\emph{relates} the ICP thermal state to the DM halo development stage, in the
form of an inverse correlation between the outer entropy slopes $a_R$ and the
halo concentrations $c$. While the outer ICP thermodynamical age is signaled
by the former, the DM dynamical age is marked by the latter, specifically in
terms of $c(z)\approx 3.5\, (1+z_t)/(1+z)$ increasing from the formation
$z_t$ to the observation redshift $z$.

We interpret such a correlation as follows. The LEs are associated with
high-$c$ halos, \emph{old} enough as to allow the ICP to be affected by deep
radiative erosion of their central entropy (producing low $k_c$ values) and
by reduced entropy production in the outskirts (shallow $a_R$ or low $k_2$);
the latter effect inescapably depends on large-scale
cosmogonical/cosmological evolution, and at given $c$ is more pronounced at
low $z$ when reduced accretion is most effective. Conversely, the HEs (and
$\mathrm{\widetilde{HE}}$s) are associated with \emph{young} halos of low
$c\la 4$. We stress that such central and outer ICP thermal evolutions are
independently driven at far apart locations by different processes; what they
have in common, though, is their largely parallel progress over comparable
timescales of several Gyrs.

On the other hand, a reasonable amount of variance in central entropy and in
outer bending may produce some intermediate instances; one such case is
constituted by A1689 at $z\approx 0.18$, with its still rather high $k_c$
level and intermediate values of its outer entropy slope. As a matter of
fact, considerable variance around the average picture will be caused by the
well-known scatter in the birth and development of cosmic structures; this
affects both the halo collapse redshifts $z_t$ (e.g., Bullock et al. 2001,
Wechsler et al. 2006, Klypin et al. 2011), and the subsequent merging
histories (e.g., McCarthy et al. 2007, Fakhouri et al. 2010). In particular,
A1689 with its mass $M\approx 1.3\times 10^{15}\, M_\odot$ constitutes a
well-studied case of high $z_t\approx 2.5$ as inferred from its high
concentration $c\ga 10$ (see Appendix A; also Broadhurst et al. 2008, Lapi \&
Cavaliere 2009b).

Another source of variance is constituted by the cluster environment; in
particular, adjoining filaments with contrasts $\delta\rho/\rho\sim 5$ will
enhance diffuse accretion in a rich ambient like a supercluster, so as to
delay weakening of shocks and onset of turbulence. This may be the case with
one sector out of four in A1689 (see Kawaharada et al. 2010, Molnar et al.
2010) and with A2199 (see Rines et al. 2002), implying the spherically
averaged values of $a$ to be higher than in standard LE clusters; the
opposite holds true for cluster sectors facing voids.

Additional variance might arise by cold subclumps in sectors of the nearby
Virgo and Perseus clusters; this would bias high the surface brightness and
the apparent baryonic fraction (see Ettori et al. 1998 vs. Simionescu et al.
2011, Urban et al. 2011). On the other hand, such features do not appear to
affect most of the clusters collected in Table~1, including instances with
steep temperature decline and flat entropy distributions like A1795; as for
the latter, our Supermodel yields an outer baryonic fraction bounded by
$0.14$. Similar values have been recently inferred from aimed \textsl{Suzaku}
observations of A2142 (see Akamatsu et al. 2011).

In LE clusters we expect \emph{outer} turbulence related to compressive modes
to develop under the drive of kinetic energy increasingly seeped through
weakening shocks (Cavaliere et al. 2011). \emph{Inner} turbulence, on the
other hand, is likely stirred in HE clusters by shear motions associated with
the mergers' wakes (e.g., Iapichino et al. 2011). These motions are widely
held to accelerate electrons in situ up to Lorentz factors $\gamma\sim 10^4$;
the electrons energize strong radiohalos by their synchrotron radiation in
cluster-wide magnetic fields of a few $\mu$Gs, with electron lifetimes under
$1$ Gyr (see Ferrari et al. 2008, Feretti et al. 2011, Brunetti 2011), shorter than the
thermal cooling times around $5$ Gyr for the center of an HE cluster. On the
other hand, in the process of cooling toward an LE state with levels $k_c\la
50$ keV cm$^2$ the core becomes sufficiently cold as to be sensitive even to
lesser mergers. Then temperature wiggles and radiohalos may form together,
but the latter will fade much sooner than the former can be eroded away (see
Buote 2002, Brunetti et al. 2009; also Rossetti et al. 2011); as a result, we
expect more $\mathrm{\widetilde{HE}}$ clusters than radiohalos.

In this paper, we have shown how entropy offers a key to detailed ICP
profiles, and a handle to physically relate the ICP state to the DM's (see
\S~5.1). In the cases we have analyzed to now, we have identified two main
cluster populations, HE and LE (see \S~5.2). We have found the former to
feature concentrations $c\approx 3-5$ associated with a slow outer decline of
$T(r)$ from a central plateau, a flat central and a steep outer brightness;
one variant of this pattern is due to the ICP being rejuvenated by mergers,
leading to the $\mathrm{\widetilde{HE}}$ subclass. The other main population
is constituted by the LE clusters. We have found these to feature: higher
concentrations $c\ga 6$, associated with a central brightness spike and low
but not vanishing central temperatures; and a steep outer decline of $T(r)$
from the inner peak, with a considerable contribution of turbulent support to
equilibrium. Such a pattern is generally sharpened toward low $z$ (see \S~6),
and implies low outer SZ signals. Finally, our picture leads us to expect a
main evolutionary sequence proceeding from HE to LE clusters. The above
classification and time developments combine into our cluster Grand Design.

In summary, in the articulated ensemble of galaxy clusters, the entropy-based
framework provided by our Grand Design offers a thread toward understanding
their basic astrophysics. Specifically, from the X-ray vantage point we
interpret the correlations between ICP and DM parameters in terms of
synchronization of the central and outer entropy demises, over timescales of
several Gyrs. Within such a context, variance may be introduced by diverse
large-scale environments adjacent to the outskirts, and possibly by
multi-phase conditions at the center. Such a variance may blur the
synchronized developments, and originate instances \emph{intermediate}
between our two main classes. We have identified one such ensemble, the
$\mathrm{\widetilde{HE}}$ clusters observed at relatively low $z$. At the
other end $z>0.5$, our Grand Design raises a specific issue concerning any
clusters where central cooling is already advanced (possibly requiring a
multi-phase ICP), while entropy production is still high in briskly
developing outskirts (see also \S~6). Observations of such objects at the
current frontier of cluster astrophysics will constitute a challenging but
rewarding aim.

\begin{acknowledgements}
The work has been supported in part by ASI/INAF agreement n. ASI-INAF
1/016/07/0. We thank our referee for stimulating us to include quantitative
correlations, and to substantially improve our presentation. We also
acknowledge useful discussions with A. Balbi, and with S. Ettori, P.
Mazzotta, S. Molendi, and P. Rosati in the context of the meeting `A New
Generation of Galaxy Clusters Surveys', July 2011 at the Sexten Center for
Astrophysics. AL thanks SISSA for warm hospitality.
\end{acknowledgements}

\clearpage

\begin{appendix}

\section{Cluster buildup}

In this Appendix we collect for the reader's convenience some \emph{basics}
of cluster formation, that are used throughout the main text to understand
the entropy distribution in the ICP in connection with the halo development
stages. Here our thrust will be to relate these stages to the shape of the
initial density perturbation; on basic grounds, such times scale as $t_{\rm
coll}\propto (G\, \rho)^{-1/2}$.

In detail, the perturbation shape is conveniently parameterized as $\delta
M/M\propto M^{-\epsilon}$, that may be considered a piecewise approximation
to a realistically bell-shaped cold DM perturbation. Here $\delta M$
represents the mass excess within a shell at the initial comoving radius
$r_i\propto M^{1/3}$ enclosing a mass $M$ at background density. Such a shell
will progressively detach from the Hubble flow, reach a maximum `turnaround'
radius $R_{\rm ta}\propto r_i/(\delta M/M)\propto M^{\epsilon+1/3}$, and
collapse back under local gravity to a virialization radius $R\approx R_{\rm
ta}/2$.

The virialization occurs when $\delta M/M$ attains the critical threshold
$1.69\, D^{-1}(t)$ in terms of the linear growth factor $D(t)$ depending on
the cosmic time $t$. So the shape parameter $\epsilon$ also governs the mass
buildup after $M(t)\propto D^{1/\epsilon}(t)\propto t^{d/\epsilon}$, where in
the standard cosmology $D(t)\propto t^d$ applies with $d$ lowering from $2/3$
to $1/2$ as $z$ decreases from above $1$ to below $0.5$. The corresponding
collapse time reads $t_{\rm coll}\equiv M/\dot M=\epsilon\, t/d$ for the
shell surrounding the mass $M$. Here $\epsilon$ marks the cosmogonic effect
of the perturbation tapering shape; on the other hand, $d$ marks the effects
of cosmology at large thinning out the background, and delaying collapse when
$d$ approaches $1/2$. In many relations to follow what matters for the
effective degree of halo development will be the combined index
$\epsilon/d=t_{\rm coll}/t$; values $\epsilon/d\la 1$ apply to the fast
collapse of the perturbation bulk, while during the slow outskirts
development the accretion rate peters out corresponding to values
$\epsilon/d\ga 1$. Note that the transition between the two regimes at $z_t$
corresponds to $\epsilon/d=1$, i.e., to the collapse time matching the Hubble
expansion timescale as per definition; of course, this agrees with the
transition epoch recognized in state-of-the-art $N-$body simulations and
semianalytic computations (see Zhao et al. 2003, Diemand et al. 2007,
Fakhouri et al. 2010, Genel et al. 2010, Wang et al. 2011). At $z_t$, the
halo concentration takes on the value $c\approx 3.5$ (with minor mass
dependence, see Prada et al. 2011), and grows afterwards as $c(z)\approx
3.5\,(1+z_t)/(1+z)$.

While the halo develops and the concentration increases, entropy is produced
by shocks in the contained ICP both at the center and at the boundary. As to
the latter shocks, the driver is constituted by the external gas inflowing
under the pull of the outer gravitational potential well; the inflow varies
during cluster development, to the effects discussed below.

\subsection{Decreasing potential drops}

The potential drop from the turnaround $R_{\rm ta}$ to the shock radius $R_s$
reads
\begin{equation}
\Delta\Phi=-\int_{R_{\rm ta}}^{R_s}{\rm d}r~{G\,\delta M\over r^2}~.
\end{equation}
As the integrand behaves like $\delta M/r^2\propto M^{1-\epsilon}/r^2\propto
r^{1-3\epsilon}$, one finds
\begin{equation}
\Delta\phi={1-\left(R_s/R_{\rm ta}\right)^{3\epsilon-2}\over 3\epsilon-2}~,
\end{equation}
where $\Delta\phi\equiv \Delta\Phi/v_{R}^2$ is for the drop normalized to the
circular velocity scale $v_{R}^2=G\, M(<R)/R$, in fact at radius $R_s$ of the
boundary shock. The potential drop as a function of $\epsilon$ is illustrated
in Fig.~A1.

\subsection{Outgrowing shock positions}

The position $R_s$ of the shock may be determined from the scaling laws
\begin{equation}
v_1^2\propto {M\over R_s}\, \Delta\phi~~~~~\dot M\propto \rho\,v_1\, R_s^2;
\end{equation}
here $M\propto \rho\, R_s^3$ is the overall mass within $R_s$, $v_1$ is the
infall velocity in the cluster frame, $\rho$ the background density (we have
assumed $n_1\propto \rho$ and $m\propto M$), and $\Delta\phi$ the
adimensional potential drop described above.

Combining the scaling laws yields
\begin{equation}
R_s\propto {M\over \dot M^{2/3}}\, (\Delta\phi)^{1/3}\propto
\left({\epsilon\over d}\right)^{2/3}\, (\Delta\phi)^{1/3}\,
t^{(d/\epsilon+2)/3}~,
\end{equation}
that, when normalized to the turnaround radius $R_{\rm ta}\propto
M^{\epsilon+1/3}$, may be written in the form
\begin{equation}
{R_s\over R_{\rm ta}}\propto \left({\epsilon\over d}\right)^{2/3}\,
(\Delta\phi)^{1/3}\,t^{-d+2/3}~;
\end{equation}
as $d$ takes on values within the narrow range $2/3-1/2$, the explicit time
dependence is very weak and may be neglected.

Using the expression of $\Delta\phi$ derived in the previous \S~A1, the
following equation for $x\equiv R_s/R_{\rm ta}$ obtains
\begin{equation}
{x^3\over 1-x^{3\epsilon-2}}=\mathcal{N}\,{(\epsilon/d)^2\over 3\epsilon-2}~;
\end{equation}
the normalization factor $\mathcal{N}$ is set on requiring that at the
transition $\epsilon=d$ the potential drop takes on the value
$\Delta\phi\approx 0.57$ corresponding to $a\approx 1.1$ after Eq.~(4) of the
main text. This yields a shock radius $R_s\la R\approx R_{\rm ta}/2$ close to
the virial boundary during the early stages of cluster buildup that involve
high accretion rates, corresponding to $\epsilon\la 1$.

The position of the shock radius and the corresponding values of the
potential drop are illustrated in Fig.~A1. During the early collapse when
$\epsilon\ll 1$ applies we find the approximations $x\propto
\epsilon^{2/5}\rightarrow 0$ and $\Delta\phi\propto \epsilon^{-4/5}$ to hold.
At the other end, during the late outskirts development when $\epsilon\gg 1$
applies we obtain $x\rightarrow 1$ and $\Delta\phi\propto\epsilon^{-2}$; thus
the shock positions outgrows the virial boundary to approach the turnaround
in the late development stage.

\subsection{Decreasing infall speeds and shock strengths}

From the scaling laws Eqs.~(A3) we also derive an expression for the infall
velocity (in the cluster frame)
\begin{equation}
v_1 \propto \dot{M}^{1/3}\,(\Delta\phi)^{1/3}\propto \left({\epsilon\over
d}\right)^{-1/3}\,(\Delta\phi)^{1/3}\, t^{(d/\epsilon-1)/3}~.
\end{equation}
This should be compared with the scaling $c_s\equiv (5\,k_BT_1/3\,\mu
m_p)^{1/2}\propto \rho^{1/3}$ of the sound speed in the preshock gas; from
Eqs.~(A3) we obtain $\rho\propto (\dot M/M)^2\, (\Delta\phi)^{-1}$ for the
density at the cluster edge, to yield
\begin{equation}
c_s \propto {\dot{M}^{2/3}\over M^{2/3}}\,(\Delta\phi)^{-1}\propto
\left({\epsilon\over d}\right)^{-2/3}\,(\Delta\phi)^{-1/3}\, t^{-2/3}~.
\end{equation}
The ratio of the two quantities reads
\begin{equation}
{v_1\over c_s} \propto \left({\epsilon\over
d}\right)^{1/3}\,(\Delta\phi)^{2/3}\, t^{(d/\epsilon+1)/3}~,
\end{equation}
and is seen to scale as $\epsilon^{-4/5}\, t^{d/3\epsilon}$ for $\epsilon\ll
1$, and as $\epsilon^{-1}\, t^{1/3}$ for $\epsilon\gg 1$. In other words,
strong shocks with $v_1\gg c_s$ take place during the early collapse of the
cluster body, whilst during the late development of the outskirts the shocks
weaken and $v_1\ll c_s$ applies.

From Eq.~(A4) we compute the shock speed to be
\begin{equation}
v_s\equiv \dot R_s \propto \left({\epsilon\over d}\right)^{2/3}\,
\left({1\over 3}\,{d\over \epsilon}+{2\over
3}\right)\,(\Delta\phi)^{1/3}\,t^{(d/\epsilon-1)/3}~.
\end{equation}
Taking the ratio of the two quantities (A7) and (A10) yields the expression
\begin{equation}
{v_s\over v_1}={1\over 3}+{2\over 3}\,{\epsilon\over d}~.
\end{equation}
This takes on values around $1/3$ during the early collapse when $\epsilon\ll
1$; it grows during the late outskirts development when $\epsilon\gg 1$,
since $v_1$ vanishes while $v_s$ decreases toward its limiting value given by
the sound speed $c_s$. Note the ubiquitous appearance in the DM dynamics of
the key quantity $\Delta\phi$, which will also appear directly in the ICP
equilibrium condition.

\subsection{Weakening shocks and increasing seepage}

In the main text we discuss how the boundary shocks are to weaken as the
cluster outskirts develop; meanwhile, an increasing fraction of kinetic
energy seeps through them. Next we explain why.

The jump conditions for entropy, temperature and density across a shock front
write (Landau \& Lifshitz 1959)
\begin{eqnarray}
\nonumber {k_2\over k_1}={T_2/T_1\over (n_2/n_1)^{2/3}}~~&&~~{\rm with} \\
\\
\nonumber {T_2\over T_1}={5\over 16}\,{\tilde v_1^2\over c_s^2}+{7\over
8}-{3\over 16}\,{c_s^2\over \tilde v_1^2}~~~~&,&~~~~{n_2\over n_1}= {\tilde
v_1\over \tilde v_2} = {4\over 1+3\,c_s^2/\tilde v_1^2}~;
\end{eqnarray}
The suffix $1$ and $2$ indicate pre- and postshock values, while quantities
with a tilde refer to the shock rest frame (where the shock velocity $\tilde
v_s$ is zero by construction); in addition, $\mathcal{M}\equiv\tilde
v_1/c_s\geq 1$ is the Mach number of the accretion shock. The behavior of
these quantities as a function of $\mathcal{M}$ is illustrated in Fig.~A2.

In the cluster frame, the shock velocity $v_s$ differs from zero, and the
upstream and downstream bulk velocities are given by $\tilde
v_{1,2}=v_{1,2}+v_s$. Using Eqs.~(A10) we work out the ratio $v_2/v_1$ to be
\begin{equation}
{v_2\over v_1}={1-3\,v_s/v_1\over 4}+{3\over 4}\,{c_s^2/v_1^2\over
1+v_s/v_1}~.
\end{equation}

The above results are summarized as follows. During the early collapse of the
cluster body with $\epsilon\ll 1$, \emph{strong} shocks with $v_1/c_s\gg 1$
and $v_s\simeq v_1/3$ hold; these imply high postshock temperatures $k_B
T_2\simeq \mu m_p\, v_1^2/3\propto \epsilon^{-6/5}$ and low bulk postshock
velocities $v_2\propto \epsilon^{3/5}\simeq 0$. On the other hand, during the
late development of the cluster outskirts with $\epsilon\gg 1$, \emph{weak}
shocks with $v_1/c_s\ll 1$ and $v_s\simeq c_s$ occur, to yield low $T_2\simeq
T_1$ and $v_2\simeq v_1\propto \epsilon^{-1}$.

Thus as the cluster buildup progresses from bulk collapse to outskirts
development, at the boundary the thermal postshock energy $k_B T_2$
monotonically \emph{decreases}, but the bulk energy $v_2^2$ seeping through
the shock to drive turbulence first increases up to a \emph{maximum};
eventually, however, it decreases when the accretion becomes transonic.

\subsection{Specific clusters}

In the way of numerical values, the halo of a typical HE cluster observed at
$z\approx 0.1$ collapsed at $z_t\approx 0.5$, developing a concentration
$c\approx 4$. On the other hand, the halo of a typical LE cluster observed at
$z\approx 0.1$ collapsed at $z_t\approx 1$; during the evolution $\epsilon$
increases from initial values close to $0.3$ to values around $0.6$ at
$z\approx 0.3$, and to $1$ on moving to $z\approx 0.1$; meanwhile, its
concentration increases from initial values around $3.5$ to values around $5$
at $z\approx 0.3$, and around $7$ at $z\approx 0.1$.

In parallel, for a relaxed, long-lived LE cluster the prevailing Mach numbers
are to decline from $\mathcal{M}^2\approx 10$ to $3$ and then toward $1$ at
low $z$, while the ratio $v_2^2/v_1^2$ increases from $10\%$ to $30\%$ with a
correspondingly high outer turbulence levels, and then decreases again toward
a few percent.

Finally, one borderline instance is provided by the LE cluster A1689 at
$z\approx 0.18$; its high concentration $c\approx 10$ implies the collapse
redshift $z_t\approx 2.5$, particularly high for its mass $M\approx 1.3\times
10^{15}\, M_\odot$. The other borderline instance is provided by the HE
cluster A2218 again at $z\approx 0.18$; its concentration $c\approx 5$
implies the collapse redshift $z_t\approx 0.7$.

\section{The ICP Supermodel}

Here we recap the basics of our ICP Supermodel introduced in \S~3 (see
Cavaliere et al. 2009, Fusco-Femiano et al. 2009, Lapi et al. 2010). The
robust snapshots it provides guide our classification in cluster classes, and
establish relationships between them to constitute the cluster Grand Design.

In equilibrium conditions, the DM gravitational pull is withstood by the
gradient of the thermal ICP pressure $p$, to yield
\begin{equation}
-{G\, M(<r) \over r^2}={1\over m_p\, n(r)}\, {{\rm d}p(r)\over {\rm d}r}=-{5\,
k^{3/5}(r)\over 2\, \mu m_p}\, {{\rm d}\over {\rm d}r}\left[{k_B T(r)\over
k^{3/5}(r)}\right]~.
\end{equation}
To begin with, in the second equality we have considered only thermal
pressure $p=n\, k_B T/\mu$, expressed in terms of the specific entropy
$k\equiv k_B T/n^{2/3}$; in the main text the entropy distribution $k(r)$ is
related to definite physical processes: it is conserved by adiabatic
compressions, produced at shock fronts, eroded by radiative cooling
(nonthermal, turbulent support is dealt with in \S~B3).

Eq.~(B1) shows how, given the potential well, the ICP disposition is set by
the entropy distribution $k(r)$. In fact, it constitutes a first order linear
differential equation for $T(r)$, which solves to the profile
\begin{equation}
\bar T(r)=\bar k^{3/5}(\bar r)\,\left[1+{2\over 5}\, b_R\,\int_{\bar r}^1{{\rm
d}\bar r'\over \bar r'}\, \bar v_c^2(\bar r')\, \bar k^{-3/5}(\bar r')\right]~,
\end{equation}
while the self-consistent density profile follows $\bar n(\bar r)=[\bar
T(\bar r)/\bar k(\bar r)]^{3/2}$. Here variables with a bar are normalized to
their boundary values at $r=R$, while $b_R\equiv \mu m_p\, v_R^2/k_B T_2$
expresses in the solution the boundary condition $T_2$ provided by the shock
jumps.

The squared circular velocity $\bar v_c^2(\bar r)\equiv \bar M(<\bar r)/\bar
r$ depends on the DM mass distribution. For the latter, we use our
$\alpha$-profiles derived from the Jeans equation (see Lapi \& Cavaliere
2009a, 2009b, 2011); the corresponding density profile reads
\begin{equation}
\bar \rho(\bar r)=\bar r^{-\tau}\, \left[{1+w\, c^\eta\over 1+w\,(c\,\bar
r)^\eta}\right]^\xi~,
\end{equation}
where $\tau\approx 0.76$, $\eta\approx 0.58$, $\xi\approx 4.56$, and
$w=-(2-\tau)/(2-\tau-\eta\xi)\approx 0.88$ are constants with values suitable
for rich clusters. Note that the standard NFW profile corresponds instead to
$\tau=1$, $\eta=0$, $\xi=2$, and $w=1$. It is seen how the density profile is
modulated by $c$ (the standard concentration parameter of the DM halo defined
in \S~1 of the main text), and particularly so in the outskirts. \textsl{IDL}
and \textsl{FORTRAN} algorithms to implement the above equations can be found
at the URL \texttt{http://people.sissa.it/$\sim$lapi/Supermodel/}.

We stress that the standard models for the ICP distribution constitute useful
approximations to the Supermodel results (e.g., Cavaliere \& Fusco-Femiano
1976, Sarazin 1988); specifically, the isothermal $\beta-$model applies in
central regions with high central level $k_c$, while the polytropic model
$k(r)\propto n^{\Gamma-5/3}(r)$ with index $\Gamma\approx 1.2$ applies in
outer regions where $n(r)$ drops quickly and the temperature $T(r)$ undergoes
a mild decline.

In a nutshell, low entropy levels throughout the clusters allow the ICP
thermal velocity dispersion $T(r)$ to passively \emph{mirror} the profile of
DM velocity dispersion's $\sigma^2(r)$, as to share its radial run with a
midle peak and a decline on both sides (as shown by Cavaliere et al. 2009,
and confirmed by Hansen et al. 2010). This is because the DM and the passive
ICP (in the absence of entropy additions) settle to a comparable equilibrium
within the \emph{common} gravitational potential well. Conversely, high
entropy levels at the center cause the ICP to resist gravitational
compressions, and $T(r)$ to maximally \emph{depart} from $\sigma^2(r)$, so as
to feature a monotonic increase inwards to a central plateau.

\subsection{Entropy distributions implemented}

Here we describe the radial entropy distributions $k(r)$ implemented in the
Supermodel, following the physical motivations given in \S~2 of the main
text. As also discussed there, which of the distributions is convenient to
try first may be decided a priori from a quick look to the inner temperatures
observed in X rays.

For HE clusters (whose inner temperature profile is flat) we adopt the
distribution
\begin{equation}
\bar k(\bar r)=\bar k_c+(1-\bar k_c)\, \bar r^a~;
\end{equation}
this renders an outward rise with uniform slope $a$ from a central level
$\bar k_c$ (see Fig.~1, red solid line).

For $\mathrm{\widetilde{HE}}$ clusters (whose inner temperature profile shows
sharp wiggles) we adopt the distribution
\begin{eqnarray}
\bar k(\bar r)=\left\{
\begin{array}{ll}
\bar k_c~~~~~&{\rm for}~~\bar r\leq \bar r_f,\\
\\
\bar k_c+(1-\bar k_c)\, \left[{(\bar r-\bar r_f)/(1-\bar
r_f)}\right]^a~~~~~&{\rm for}~~\bar r> \bar r_f~.
\end{array}
\right.
\end{eqnarray}
This represents a floor with level $\bar k_c$ extending out to a radius
$r_f$, followed by an outward rise with uniform slope $a$ (see Fig.~1, orange
dashed line).

For LE clusters (whose inner temperature profiles show an outward rise up to
a maximum and then a decline) we implement the distribution
\begin{eqnarray}
\bar k(\bar r)=\left\{
\begin{array}{ll}
\bar k_b\, \left({\bar r/ \bar r_b}\right)^a~~~~~&{\rm for}~~\bar r\leq \bar r_b,\\
\\
\bar r^{a+a'}\, e^{a'\,(1-\bar r)/\bar r_b}~~~~~&{\rm for}~~\bar r> \bar r_b~,
\end{array}
\right.
\end{eqnarray}
where $\bar k_b\equiv \bar r_b^{a+a'}\,e^{a'\,(1-\bar r_b)/\bar r_b}$ applies
to ensure continuity at $r=r_b$. This renders an outwards rise with uniform
slope $a$ out to a radius $r_b$, and then a progressive bending follows
with a linear decline of the slope, with the gradient $a'\equiv (a-a_R)\,
\bar r_b/(-\bar r_b+1)$ down to the boundary value $a_R$ (see Fig.~1, blue
dotted line).

\subsection{Observables, and parameter counting}

In connection with \S~4, we detail how from $T(r)$ and $n(r)$ we proceed to
compute the profiles of the X-ray and SZ observables. Specifically, the
emission-weighted temperature is given by
\begin{equation}
\langle T(\bar{w})\rangle =T_2\, {\int_0^{\bar r}\,
\mathrm{d}\bar{\ell}~\bar{n}^2\,\Lambda[\bar T]\,\bar{T}\over
\int_0^{\bar r}\,
\mathrm{d}\bar{\ell}~\bar{n}^2\,\Lambda[\bar T]}~,
\end{equation}
where $\bar r = \sqrt{1-\bar{w}^2}$ is expressed in terms of the projected
radius $\bar{w}$; here $\Lambda[T]$ is the cooling function, with a typical
dependence $\Lambda\propto T^{1/2}$ for hot clusters with average $k_B T\ga
2$ keV.

The brightness distribution is given by
\begin{equation}
S(\bar{w})=n_2^2\, R\,\int_0^{\bar r}\,
\mathrm{d}\bar{\ell}~\bar{n}^2\,\Lambda[\bar{T}]\,F[E_H,E_L,T]~;
\end{equation}
where the factor $F[E_H,E_L,T]\simeq e^{-E_L/k_B T(r)}-e^{-E_H/k_B T(r)}$
takes into account specific instrumental bands $E_H-E_L$.

Fitting these expressions to the observations leads to pin down the following
parameters and scales. From the profile normalizations we determine the ICP
\emph{scales} $n_2$ and $T_2$, and the DM scale $R$ (if not independently
given by observations of galaxy dynamics, `red sequence' termination,
gravitational lensing).

From the profile \emph{shapes} one can determine not only the ICP
\emph{parameters} describing the entropy run, but also the DM concentration
parameter $c$, when not independently provided by gravitational lensing. For
HE clusters $2$ ICP parameters are needed, i.e., the central level $k_c$ and
the slope $a$; for LE clusters $3$ ICP parameters are needed, i.e., the outer
value of the slope $a_R$, the average derivative $a'$ of the slope, and the
radius $r_b$; finally, for the $\mathrm{\widetilde{HE}}$ clusters $3$
parameters are needed, i.e., the level $k_c$ and the extension $r_f$ of the
central floor, and the outer slope $a$.

A preliminary guideline as to which entropy shape is conveniently tried first
is provided by a quick look at the gross temperature run at the center and in
the outskirts, as discussed in \S~4. A posteriori, the values of the reduced
$\chi^2$ of the fits provide a final check.

An independent observable is provided by the Comptonization parameter that
marks the strength of the SZ effect; it can be expressed as
\begin{equation}
y(\bar{w})=n_2\, T_2\,R\,\int_0^{\bar
r}\,\mathrm{d}\bar{\ell}~\bar{n}\,\bar{T}~.
\end{equation}
In the near future interferometric instrumentations like ALMA (see
\texttt{http://science.nrao.edu/ alma/index.shtml}) will provide measurements
of comparable sensitive and resolution to the present X-ray instrumentations
of \emph{XMM-Newton} and \emph{Chandra}.

\subsection{Turbulent support}

In connection with \S~3.2, we explain how the Supermodel can be readily
extended to cover the ICP equilibrium when nonthermal, turbulent support
contributes adding to thermal pressure (see Cavaliere et al. 2011). The
relevant quantity is provided by the ratio $\delta(r)\equiv p_{\rm
nth}/p_{\rm th}$ of turbulent to thermal pressure or, equivalently, by the
ratio $\delta/(1+\delta)$ of turbulent to total pressure $p_{\rm tot}=p_{\rm
th}\, (1+\delta)$.

We expect onset of turbulence in the outskirts of relaxed LE clusters (cf.
Eq.~A13), where weakening accretion shocks leave over an appreciable bulk
energy to drive turbulent motions with maximal amplitude $\delta_R\approx
(v_2/v_1)^2$ up to $30-40\%$ at the virial radius. In fact, these motions
start up with a coherence length $L\sim R/2$ set by the pressure scale height
or by shock segmentation enforced by the adjoining filamentary structure, and
then fragment into a dispersive cascade over the `inertial range' to sizes
$\ell$ where dissipation begins. In the ICP context the dissipation scale
writes $\ell\sim (c_2/\tilde{v})^{3/4}\, \lambda_{\rm pp}\, (L/\lambda_{\rm
pp})^{1/4}$ in terms of the ion collisional mean free path $\lambda_{\rm
pp}\approx 50$ kpc and of the ratio $\tilde{v}/c_2$ of the turbulent rms
speed to the sound's. For \emph{subsonic} turbulence with
$\tilde{v}/c_2\lesssim 1/3$, the relevant scale $\ell$ exceeds somewhat
$\lambda_{\rm pp}\sim 100$ kpc. This behavior may be rendered in terms of the
simple functional shape
\begin{equation}
\delta(r)=\delta_R\, e^{-(R-r)^2/\ell^2}~,
\end{equation}
which decays on the \emph{scale} $\ell$ inward of a round plateau, a smoothed
out representation of the inertial range. This provides the gradient of the
turbulent pressure.

In fact, the equation of hydrostatic equilibrium in the presence of turbulent
support will contain the total pressure in the general form $p_{\rm th}(r)\,
[1+\delta(r)]$, while the thermal component is still expressed as $p_{\rm
th}(r)\propto k(r)\, n^{5/3}(r)$. On noting that $p_{\rm tot}=p_{\rm th}\,
(1+\delta)=n\, k_B \tilde T/\mu$ with $\tilde T\equiv T\, (1+\delta)$, it is
convenient to introduce the \emph{extended} entropy
\begin{equation}
\tilde k\equiv k\, (1+\delta)~.
\end{equation}
This quantity renders the conversion of kinetic energy into random energy at
two levels, the microscopic one given by the standard $k$, and the dispersion
into turbulent `eddies' given by $k~\delta$. It is easily checked that in
these terms the solution has the same form of Eq.~(B2).

It turns out that the profiles of emission-weighted temperature are little
affected by turbulence (see Cavaliere et al. 2011); on the other hand, the
thermal SZ effect is lowered to $\tilde y\equiv y/(1+\delta)$. Finally,
including turbulence in the equation of hydrostatic equilibrium brings the
total mass reconstructed from X rays into agreement with the findings from
simulations and with that measured via gravitational lensing observations
(see Nagai et al. 2007, Lau et al. 2009, Meneghetti et al. 2010, Cavaliere et
al. 2011).

\end{appendix}
\clearpage

\clearpage
\begin{figure}
\epsscale{1}\plotone{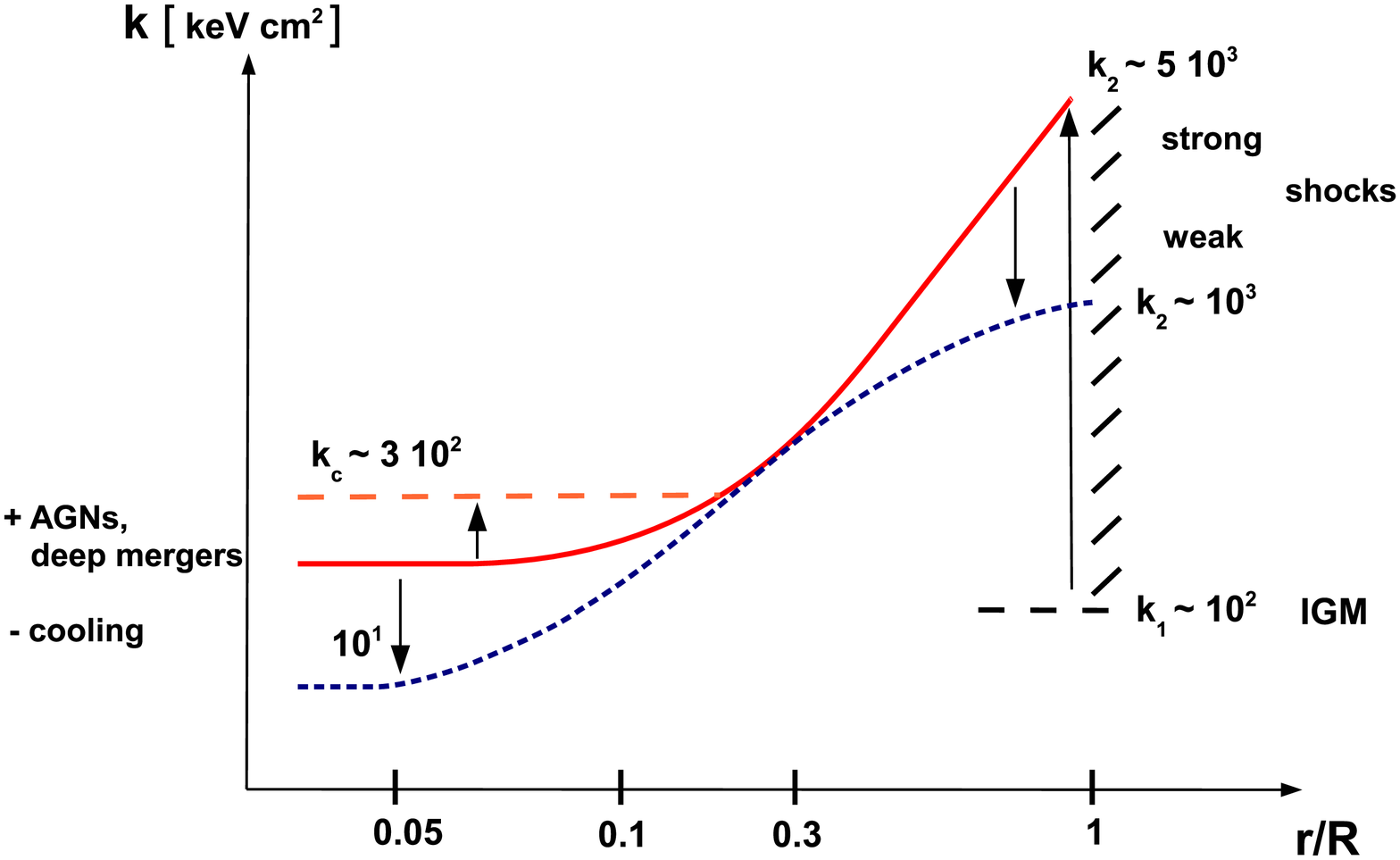}\caption{The schematics illustrates our fiducial
patterns for the ICP entropy distribution $k(r)$. In the basic pattern
(central level plus ramp; red solid line), entropy is raised at the
boundary from intergalactic values $k_1\sim 10^2$ keV cm$^2$ to high outer
levels $k_2\sim 5\times 10^3$ keV cm$^2$ by strong boundary shocks. As the
outskirts develop, the shocks weaken and the outer level lowers to $k_2\la
10^3$ keV cm$^2$; meanwhile, the central entropy is eroded by radiative
cooling down to low levels $k_c\approx 10^1$ keV cm$^2$ (blue dotted line).
On the other hand, blastwaves driven by deep mergers may enhance the central
levels up to $k_c\sim 3\times 10^2$ keV cm$^2$, spread out in the form of an
extended floor (orange dashed line).}
\end{figure}

\clearpage
\begin{figure}
\epsscale{1}\plotone{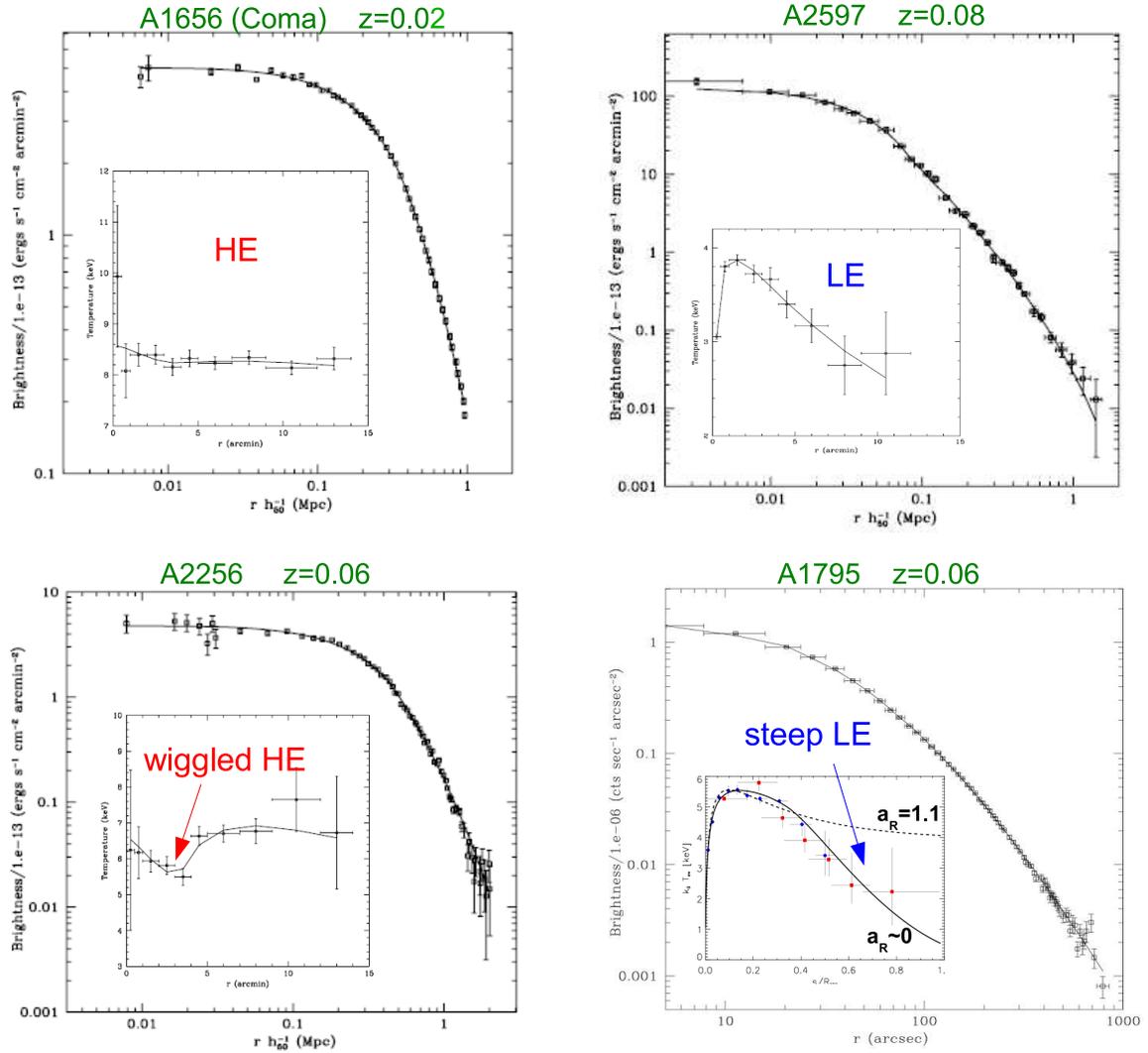}\caption{Supermodel fits to the brightness and
temperature profiles for the clusters A1656 (top left panel), A2597 (top
right), A2256 (bottom left), and A1795 (bottom right). Details are provided
in Cavaliere et al. (2009), Fusco-Femiano et al. (2009), and Lapi et al.
(2010). In the temperature panel of A1795, the Supermodel fit with a
bending entropy profile is reported as a solid line, while the fit with a
powerlaw entropy profile is reported with a dashed line.}
\end{figure}

\clearpage
\begin{figure}
\epsscale{1}\plotone{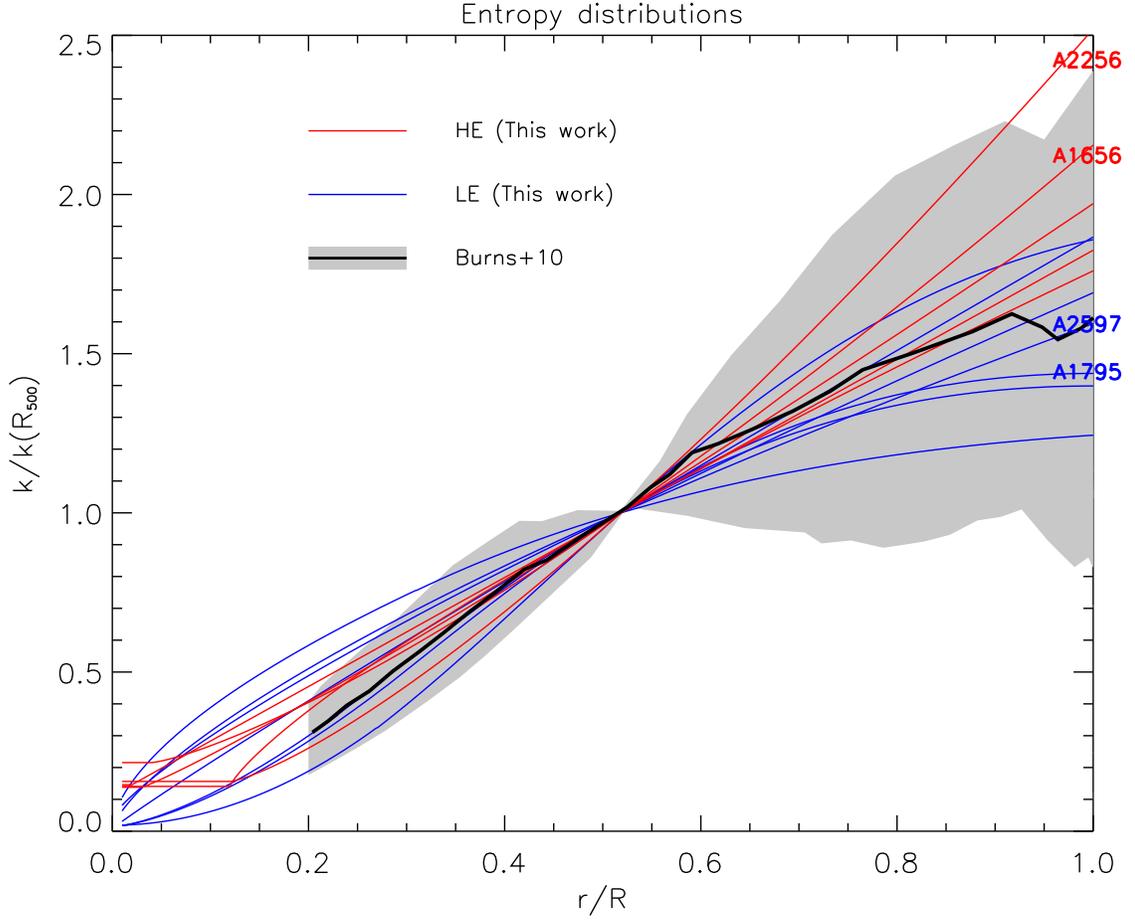}\caption{Entropy profiles normalized at
$R_{500}$ as determined with the Supermodel for the $12$ clusters listed in
Table~1; red lines refer to HE and blue lines to LE clusters. In the radial
range $r\ga 0.2\, R$ where cooling is negligible, these are overplotted to
the outcomes of the nonradiative hydro-simulations by Burns et al. (2010);
the black solid line represents the average over their sample of $24$ relaxed
clusters, with related variance illustrated by the shaded area. The four
clusters presented in Fig.~2 are labeled.}
\end{figure}

\clearpage
\begin{figure}
\epsscale{0.8}\plotone{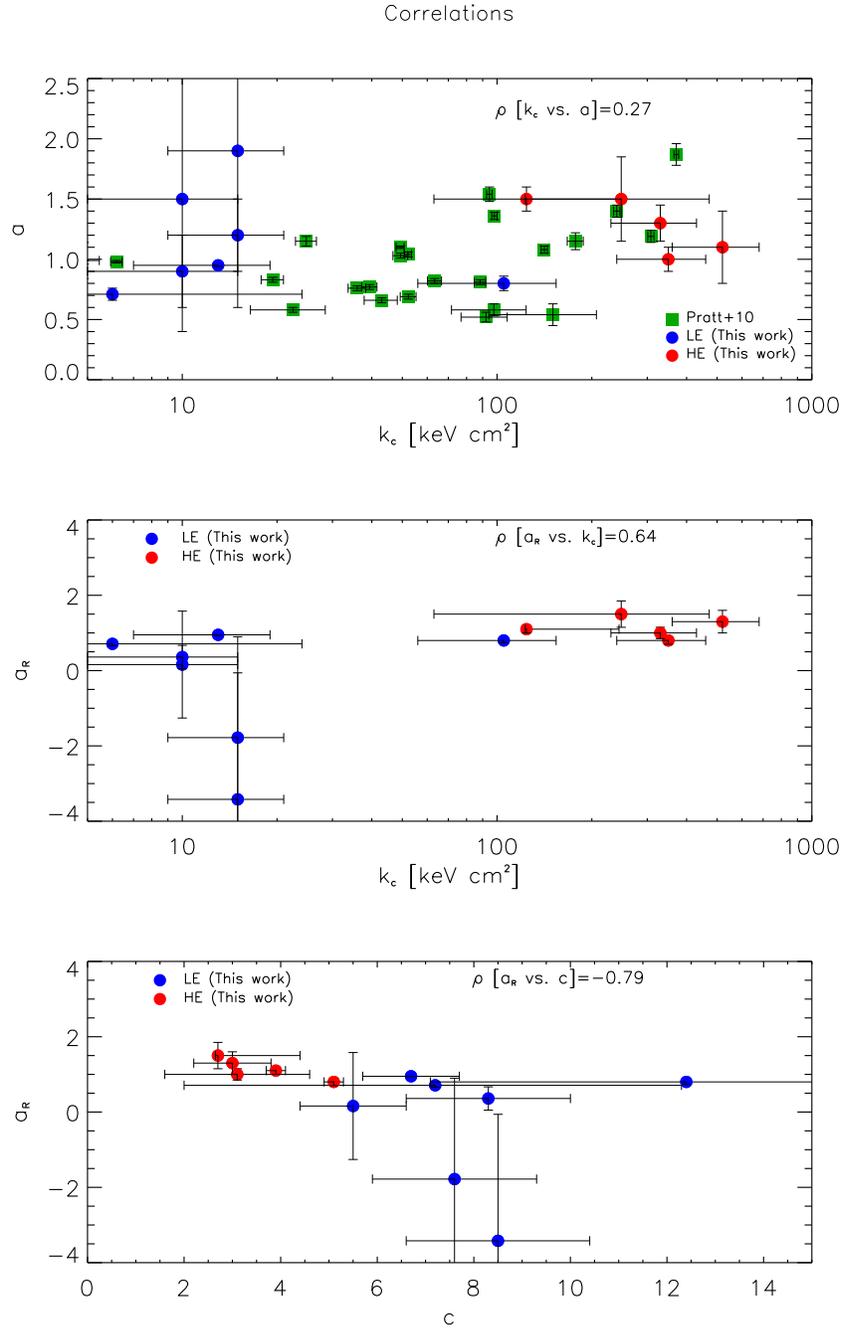}\caption{Top panel: central entropy level $k_c$
vs. entropy slope $a$ in the cluster bulk. Dots illustrate our results from
the Supermodel analysis of the $12$ clusters listed in Table~1 (red dots
refer to HEs and blue dots to LEs); squares are from the sample of $29$
relaxed clusters by Pratt et al. (2010). Middle panel: central entropy level
$k_c$ vs. outer entropy slope $a_R$; symbols are as above. Bottom
panel: the DM concentration $c$ vs. the outer ICP entropy slope $a_R$; symbols
are as above. In all panels the Spearman's rank correlation
coefficients $\rho$ for the average data values are reported.}
\end{figure}

\clearpage
\begin{figure}
\epsscale{1}\plottwo{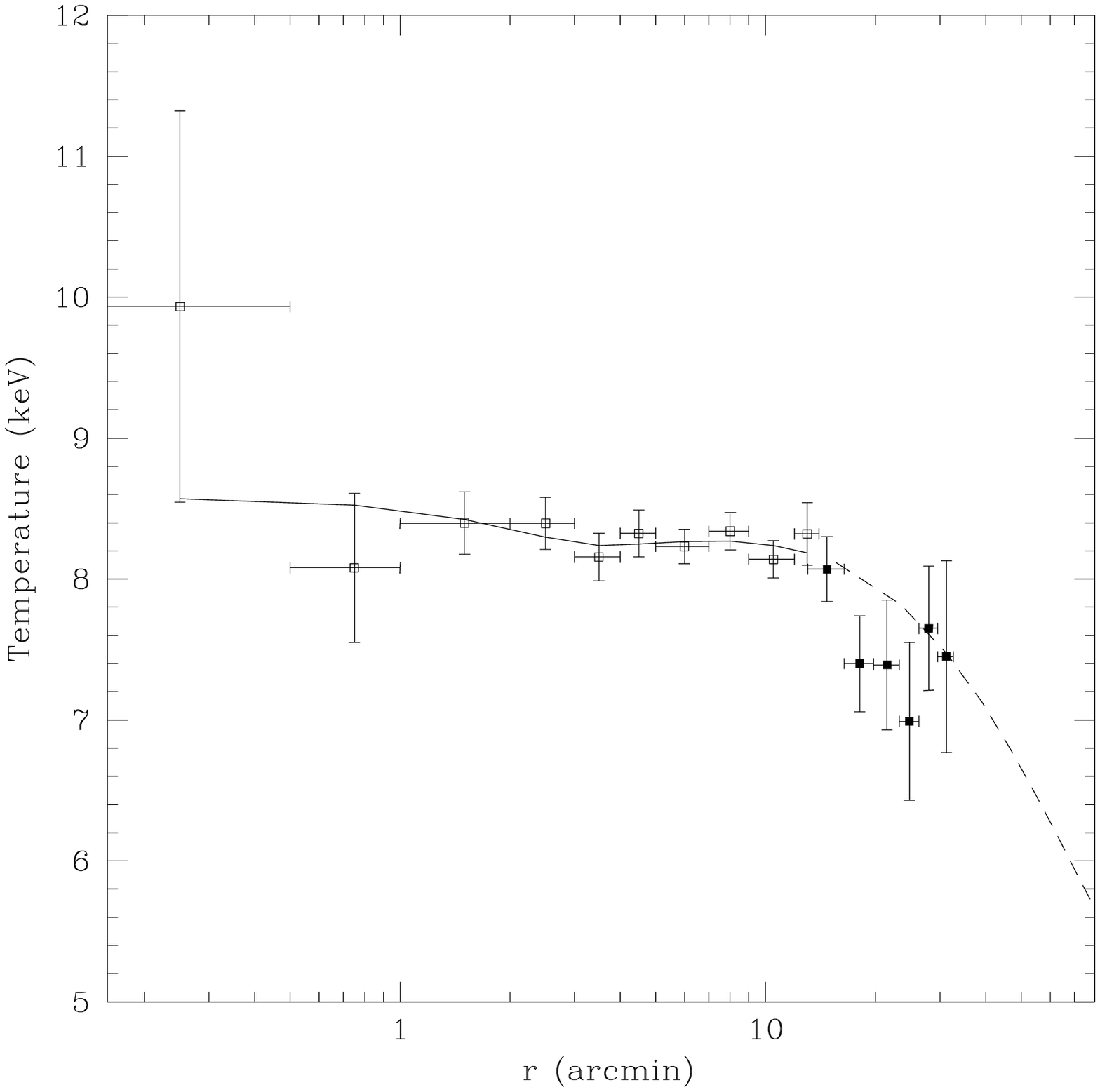}{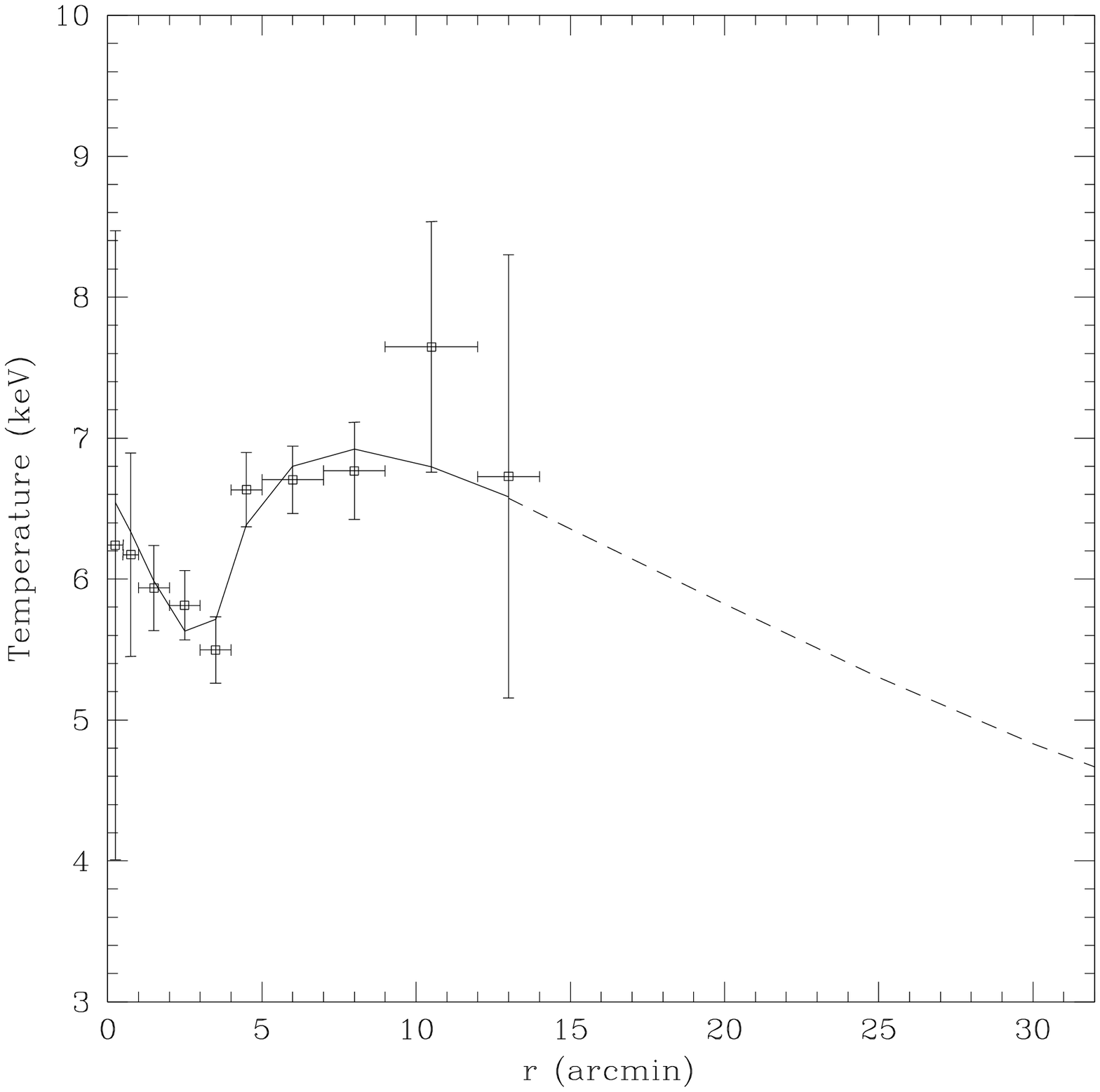}\caption{Outer temperature profiles
$T(r)$ we predict with the Supermodel from the existing inner/middle data
concerning the clusters A1656 (left panel) and A2256 (right panel); data for
the former are from Snowden et al. (2008) and Wik et al. (2009), and for the
latter from Snowden et al. (2008). The solid line represents our Supermodel
fit in the region covered by the data, while the dashed line illustrates our
prediction in the outskirts (for A1656 the fit has been performed basing
solely on the inner data by Snowden et al. 2008).}
\end{figure}

\clearpage
\begin{figure}
\epsscale{1}\plotone{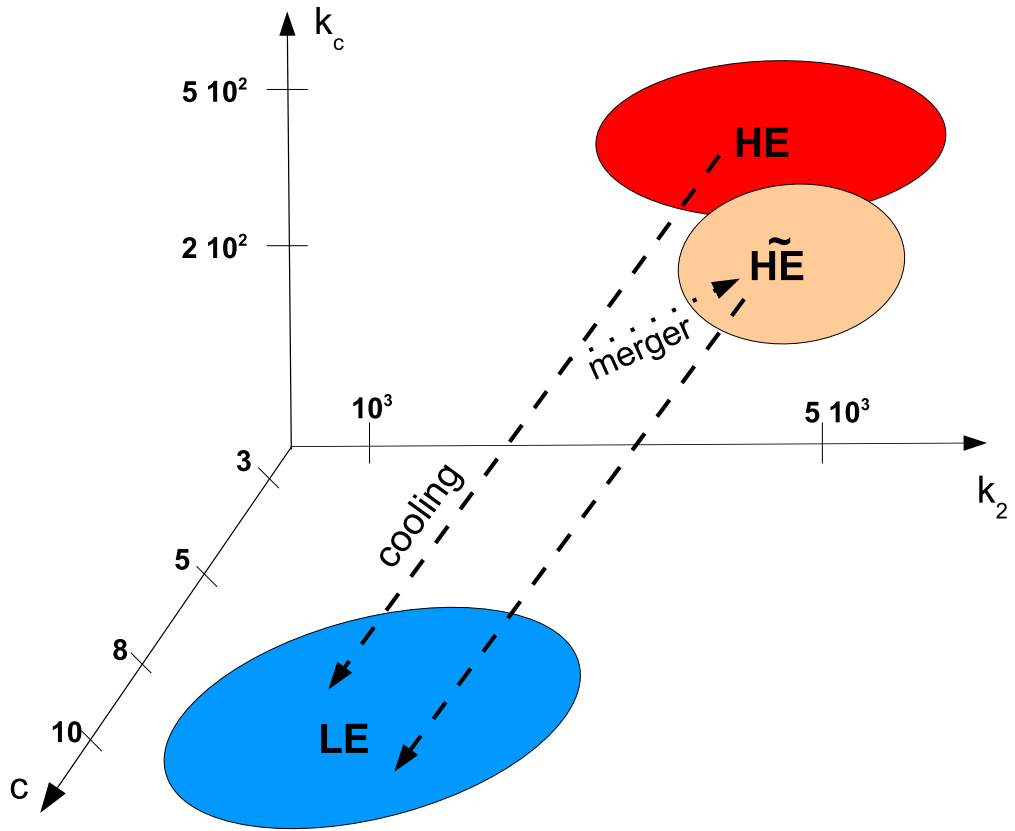}\caption{The schematics illustrates the
relationships among the cluster classes after the Grand Design.}
\end{figure}

\clearpage
\begin{figure}
\figurenum{A1}
\epsscale{0.8}\plotone{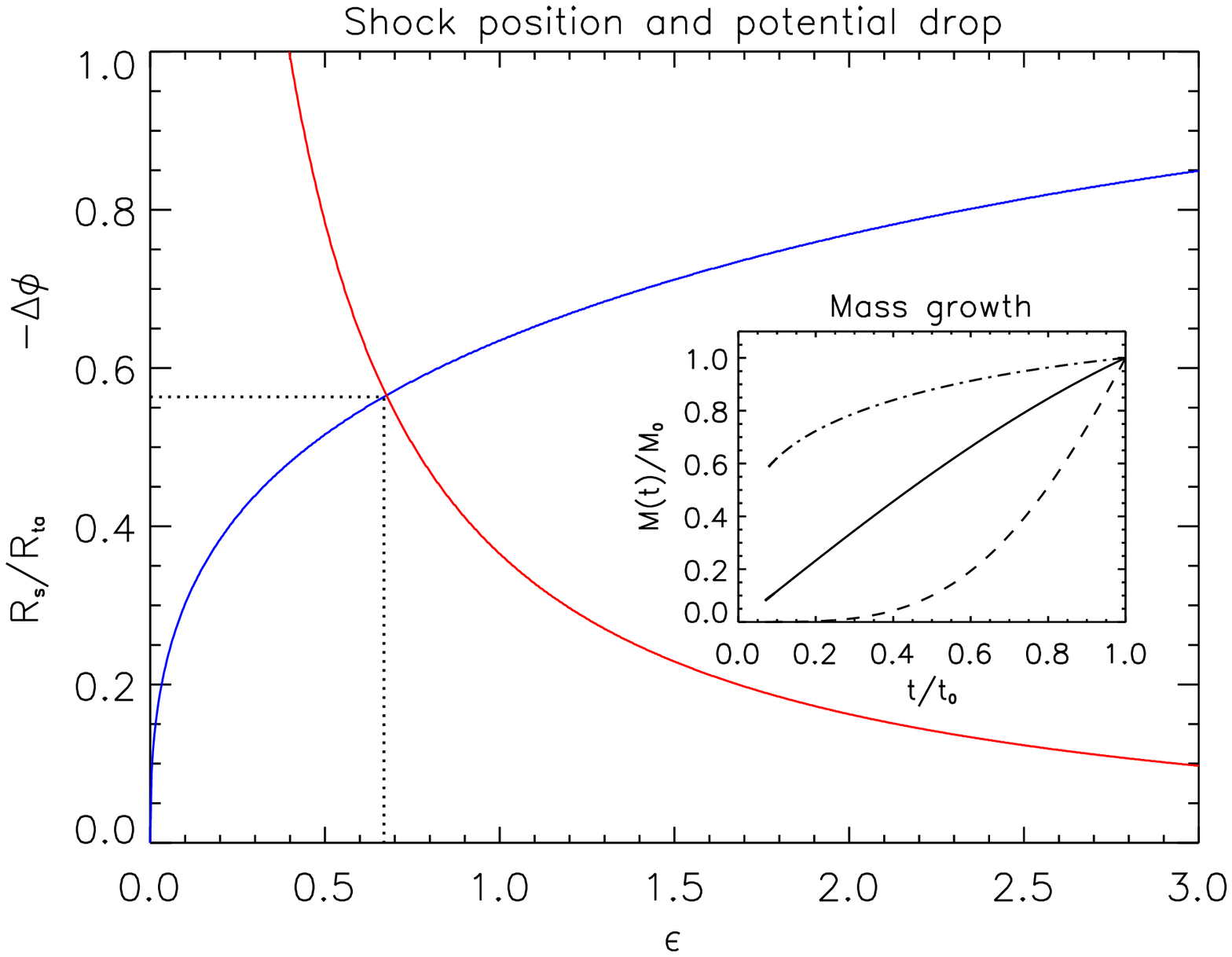}\caption{Shock position $R_s/R_{\rm ta}$
relative to the turnaround radius and potential drop $\Delta\phi$ normalized
to $v_R^2$, as a function of the parameter $\epsilon$; the inset illustrates
the mass growth $M(t)\propto t^{d/\epsilon}$ for three values of
$\epsilon=1/6$, $2/3$, and $3$, that span the range $t_{\rm
coll}/t=\epsilon/d$ going from $1/4$ to $6$.}
\end{figure}

\clearpage
\begin{figure}
\figurenum{A2}
\epsscale{0.8}\plotone{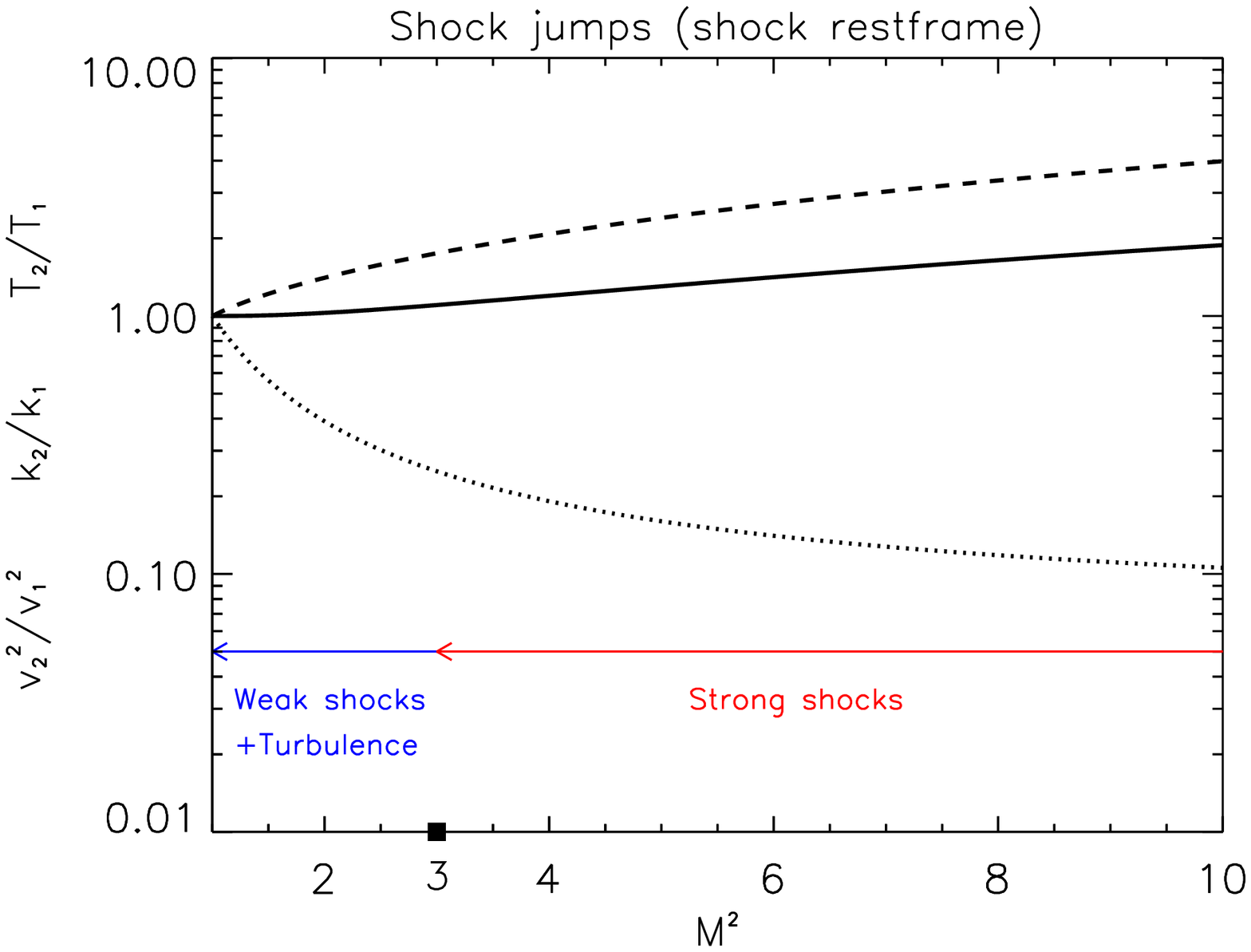}\caption{Plot of the shock jumps $k_2/k_1$,
$v^2_2/v^2_1$ and $T_2/T_1$ as a function of the squared Mach number
$\mathcal{M}^2$ in the shock restframe; note that the divide between strong
and weak shocks (associated with onset of turbulence) is around
$\mathcal{M}^2=3$. During a cluster's evolution, the outskirts condition
progresses from right to left.}
\end{figure}

\clearpage
\begin{deluxetable}{lcccccccccccccc}
\tabletypesize{}\tablecaption{A cluster library} \tablewidth{0pt}
\tablehead{\colhead{Cluster} & \colhead{Class} && $k_c$ && $r_f/R$ && $a_R$
&& $r_b/R$ && \colhead{$c$} &&\colhead{$\chi^2$} \\ & && [keV cm$^2$] &&
$[\times 10^{-2}]$ && && $[\times 10^{-2}]$ && &&}\startdata
A1795$^\dag$  & LE && $15^{+6}_{-6}$ && $-$ && $-3.43^{+3.36}_{-3.36}$ && $28^{+2}_{-2}$ && $8.5^{+1.9}_{-1.9}$ &&
0.3
(2.6)\\
PKS0745$^\dag$ & LE && $15^{+6}_{-6}$ && $-$ && $-1.78^{+2.68}_{-2.68}$ && $23^{+3}_{-3}$ && $7.6^{+1.7}_{-1.7}$ &&
1.4
(4.4)\\
A2204$^\dag$  & LE && $10^{+5}_{-5}$ && $-$ && $0.16^{+1.41}_{-1.41}$ && $31^{+7}_{-7}$ && $5.5^{+1.1}_{-1.1}$ &&
1.1
(2.1)\\
A1413$^\dag$  & LE && $10^{+5}_{-5}$ && $-$ && $0.36^{+0.31}_{-0.31}$ && $27^{+7}_{-7}$ && $8.3^{+1.7}_{-1.7}$ &&
1.2
(1.9)\\
A2597$^\ddag$  & LE && $6^{+18}_{-4}$ && $-$ && $0.71^{+0.05}_{-0.05}$ && $-$ && $7.2^{+5.0}_{-5.2}$ && 0.3\\
A2199$^\ddag$  & LE && $13^{+6}_{-6}$ && $-$ && $0.95^{+0.01}_{-0.01}$ && $-$ && $6.7^{+1.0}_{-1.0}$ && 3.1\\
A1689$^{\dag\,\ddag}$  & LE && $105^{+49}_{-49}$ && $-$ && $0.80^{+0.06}_{-0.06}$ && $-$ && $12.4^{+5.3}_{-5.3}$ &&
1.7\\
\\
A2218$^\star$  & HE && $350^{+110}_{-110}$ && $-$ && $0.8^{+0.1}_{-0.1}$ && $-$ && $5.1^{+0.2}_{-0.2}$ && 0.15\\
A399$^\star$  & HE && $330^{+100}_{-100}$ && $-$ && $1.0^{+0.1}_{-0.2}$ && $-$ && $3.1^{+1.5}_{-1.5}$ && 1.3\\
A1656$^\star$  & HE && $520^{+160}_{-160}$ && $-$ && $1.3^{+0.5}_{-0.2}$ && $-$ && $3.0^{+0.8}_{-0.8}$ && 0.7 \\
A644$^\ddag$ & $\mathrm{\widetilde{HE}}$ && $124^{+120}$ && $3^{+0.2}_{-0.2}$ && $1.1^{+0.1}_{-0.1}$ && $-$ &&
$3.9^{+0.2}_{-0.2}$ && 0.5 (3.1)\\
A2256$^{\ddag\,\star}$ & $\mathrm{\widetilde{HE}}$ && $248^{+224}_{-185}$ && $12^{+4}_{-4}$ && $1.5^{+0.4}_{-0.3}$
&& $-$
&& $2.7^{+1.7}$ && 0.9 (2.5)\\
\enddata
\tablecomments{Supermodel fits to the X-ray observables performed and/or
refined in this work (marked with a $\star$), and in the references
Fusco-Femiano et al. (2009; marked with a  $\ddag$), and Lapi et al. (2010;
marked with a  $\dag$). Dashes in the columns of $r_b/R$ ($r_f/R$) indicate
large (small), irrelevant values. The last column provides the values of the
reduced $\chi^2$ for the temperature fits, including $r_f$ or $r_b$ when
necessary (in parentheses the values obtained on using the simple powerlaw
entropy run of Eq.~2). Note that for A2218, A399, A1656 the values
$r_f=12^{+4}_{-4}\,,\, 2^{+0.1}_{-0.1}\,,\, 4^{+0.2}_{-0.2}$, respectively,
come from the centrally flat brightness profile (see discussion by
Fusco-Femiano et al. 2009).}
\end{deluxetable}

\end{document}